\documentclass[12pt,preprint]{aastex}
\lefthead{Skinner et al.}
\righthead{Sigma Orionis}

\begin{document}

\title{High-Resolution {\em Chandra} X-ray Imaging and Spectroscopy of the 
       $\sigma$ Orionis Cluster} 

\author{Stephen L. Skinner and Kimberly R. Sokal}
\affil{CASA, Univ. of Colorado, Boulder, CO 80309-0389 }

\author{David H. Cohen}
\affil{Dept. of Physics and Astronomy, Swarthmore College, 
       Swarthmore, PA 19081}

\author{Marc Gagn\'{e}}
\affil{Dept. of Geology and Astronomy, West Chester Univ., 
       West Chester, PA 19383-2130}

%%\author{Manuel G\"{u}del}
%%\affil{Paul Scherrer Inst., W\"{u}renlingen and Villigen, 
%%     CH-5232 Villigen, Switzerland}

\author{Stanley P. Owocki and Richard D. Townsend}
\affil{Bartol Research Inst., Univ. of Deleware, 
       Newark, DE  19716}

% The abstract environment prints out the receipt and acceptance dates
% if they are relevant for the journal style.  For the aasms style, they
% will print out as horizontal rules for the editorial staff to type
% on, so long as the author does not include \received and \accepted
% commands.  This should not be done, since \received and \accepted dates
% are not known to the author.
%
% Define symbol \ltsimeq
\newcommand{\ltsimeq}{\raisebox{-0.6ex}{$\,\stackrel{\raisebox{-.2ex}%
{$\textstyle<$}}{\sim}\,$}}
%
% Define symbol \gtsimeq
\newcommand{\gtsimeq}{\raisebox{-0.6ex}{$\,\stackrel{\raisebox{-.2ex}%
{$\textstyle>$}}{\sim}\,$}}

\begin{abstract}
We present results of a 90 ksec {\em Chandra} X-ray observation
of the young $\sigma$ Orionis cluster (age $\sim$3 Myr)
obtained with the High Energy Transmission Grating Spectrometer.
We use the high  resolution grating spectrum and
moderate resolution CCD spectrum of the massive central star
$\sigma$ Ori AB (O9.5V $+$ B0.5V) to  test wind shock theories of 
X-ray emission and also analyze the high spatial resolution 
zero-order ACIS-S image of the central cluster region.

{\em Chandra} detected 42 X-ray sources on the primary CCD
(ACIS-S3). All but five  have near-IR or 
optical counterparts and about one-fourth 
are variable. Notable high-mass stellar detections are
$\sigma$ Ori AB, the magnetic B star $\sigma$ Ori E, and 
the B5V binary HD 37525. Most of the other detections have  
properties consistent with lower mass K or M-type stars.
We present the first
X-ray spectrum of the unusual infrared source IRS1, located
$\approx$3$''$ north of  $\sigma$ Ori AB. Its X-ray properties
and elongated mid-IR morphology suggest it is an embedded low mass
T Tauri star  whose disk/envelope is being photo-evaporated
by  $\sigma$ Ori AB.

We focus on the radiative wind shock interpretation of the 
soft luminous X-ray emission from  $\sigma$ Ori AB, but also
consider possible alternatives including magnetically-confined 
wind shocks and colliding wind shocks. Its emission lines
show no significant asymmetries or  centroid shifts 
and are moderately broadened to HWHM $\approx$ 264  km s$^{-1}$, 
or one-fourth the terminal wind speed. 
Forbidden lines in He-like ions are formally 
undetected, implying strong UV suppression. The Mg XI triplet forms
in the wind acceleration zone within one stellar radius
above the surface. These X-ray properties are consistent
in several respects  with the predictions of radiative wind shock
theory for an optically thin wind, but explaining the narrow
line widths  presents a challenge to the theory.

\end{abstract}

% The different journals have different requirements for keywords.  The
% keywords.apj file, found on aas.org in the pubs/aastex-misc directory, 
% contains a list of keywords used with the ApJ and Letters.  These are 
% usually assigned by the editor, but authors may include them in their 
% manuscripts if they wish. 


% In the first two sections, you should notice the use of the LaTeX \cite
% command to identify citations.  The citations are tied to the
% reference list via symbolic KEYs.  We have chosen the first three
% characters of the first author's name plus the last two numeral of the
% year of publication.  The corresponding reference has a \bibitem
% command in the reference list below.
%
% Please see the AASTeX manual for a more complete discussion on how to make
% \cite-\bibitem work for you.   

\newpage
\clearpage
\section{Introduction}
Sensitive X-ray grating observations of massive OB stars 
with  {\em Chandra} and {\em XMM-Newton} during the past several 
years have provided  a wealth of new data that is improving our
understanding of the physical processes that produce their
X-ray emission.  Luminous OB stars
are not expected to have the outer convection zones needed to 
sustain internally-generated magnetic fields and solar-like
(coronal) X-ray emission. Thus, their X-ray emission has generally
been attributed to wind shocks. 

Theoretical
models have attempted to explain the X-ray emission of {\em single} 
massive OB stars in terms of X-ray emitting shocks distributed
throughout their  powerful winds that
form as a result of line-driven flow instabilities (Lucy \&
White 1980; Lucy 1982; ; Feldmeier et al. 1997;
Owocki, Castor, \& Rybicki 1988). The radiative wind shock emission 
is predicted to be soft (kT $<<$ 1 keV), in contrast to 
the harder emission (kT $\gtsimeq$ 2 keV) that has been detected 
in some massive colliding wind binaries and in a few OB stars with
strong $\sim$kG magnetic fields whose winds are 
magnetically-confined (Secs. 3.4 and 3.7.2).
X-ray grating observations provide crucial information 
on line shapes, widths, centroids, and fluxes  needed
to test wind shock models. 

Grating observations of a limited sample of O and early B-type stars 
have provided some  support for the radiative wind shock picture.
Perhaps the best support so far comes from the O4f star $\zeta$ Pup,
which has a high mass-loss rate log $\dot{M}$ $\sim$ $-$5.3 
M$_{\odot}$ yr$^{-1}$ (Pauldrach et al. 1994). Its X-ray 
emission lines are asymmetric with blueshifted centroids and are
quite broad with half-width half-maximum
values HWHM $\sim$ 900 km s$^{-1}$, equivalent to $\approx$40\%
of the terminal wind speed    
(Cassinelli et al. 2001; Kahn et al. 2001; Kramer et al. 2003). 
An analysis of a {\em Chandra} grating spectrum of the
O9 supergiant  $\zeta$ Ori by Cohen et al. (2006) also
revealed broad blueshifted asymmetric emission lines,
a different conclusion than was reached by Waldron \& Cassinelli (2001).
The blueshifted asymmetry
is predicted by radiative wind shock models of  
optically thick winds where the redward part of the line is
attenuated from receding wind material on the far side of the 
star (Owocki \& Cohen 2001).

For late O and early B-type stars with lower mass-loss rates than 
O supergiants, clear confirmation of radiative wind shock theory has been more
difficult to obtain. Grating spectra of the Orion belt eclipsing binary system
$\delta$ Ori (O9.5 II $+$ B0.5 III; 
log $\dot{M}$ $\sim$ $-$6. M$_{\odot}$ yr$^{-1}$;  
Miller et al. 2002) reveal cool plasma at kT $\sim$ 0.3 keV,
and symmetric unshifted emission lines with 
HWHM $\sim$ 430 km s$^{-1}$, or $\approx$22\% of the terminal
wind speed. The dominant cool plasma is compatible with a shock
origin and symmetric unshifted lines are predicted by  radiative
wind shock  simulations for an  optically thin wind
(MacFarlane et al. 1991; Owocki \& Cohen 2001). However, Miller et al.
(2002) raised some questions about the surprising narrowness and symmetry
of the lines in the framework of  radiative wind shock models.
An even more extreme case is the early B giant $\beta$ Cru (B0.5 III;
log $\dot{M}$ $\sim$ $-$8. M$_{\odot}$ yr$^{-1}$; Cohen et al. 2008).
It also has a low plasma temperature and symmetric unshifted lines
which are quite narrow (HWHM $\sim$ 150 km s$^{-1}$).
Cohen et al. (2008) concluded that  the narrow lines were difficult
to understand in the framework of radiative wind shock 
theory for fast winds.

Only a handful of X-ray grating observations of OB stars
have been obtained and observations of a broader sample of
objects are needed to clarify how X-ray properties vary 
with  spectral type, luminosity
class, and mass-loss parameters.
The Orion complex is particularly well-suited to observational
tests of  X-ray emission models in massive young stars because
of its relative proximity and large concentration  of very
young OB stars. It is the closest giant molecular cloud
containing young massive stars. The density of
young stars in  Orion is high  and the deep 
{\em Chandra} observation of the Orion Nebula obtained
in the COUP project detected more than 1400 X-ray sources
(Getman et al. 2005). {\em Chandra's} superb angular
resolution is clearly an advantage in deep X-ray 
observations of Orion in order to  avoid source misidentification.

A  promising target
for exploring X-ray production in young high-mass stars is
$\sigma$ Orionis AB (Table 1), the focus of this study.
It is located in the Ori OB1b association and is quite young with 
an estimated age of 3$\pm$2  Myr (Caballero 2007a and references
therein; Caballero 2008). The
{\em Hipparcos} distance is 352$^{+166}_{-85}$ pc,
which we adopt here. 
A slightly smaller value of 334$^{+25}_{-22}$ pc
was recently obtained by Caballero (2008) who also
noted that d  $\sim$ 385 pc is possible if $\sigma$ Ori AB
is a  triple system.  
For comparison, recent distance measurements of the Orion Nebula cluster
give 389$^{+24}_{-21}$ pc (Sandstrom et al. 2007),
392$^{+32}_{-32}$ pc (Jeffries 2007), 
414 $\pm$ 7 pc (Menten et al. 2007), and
391$^{+12}_{-9}$ pc (Mayne \& Naylor 2008).

$\sigma$ Ori AB is a visual binary consisting of a O9.5V primary and 
a B0.5V companion separated by  0.$''$25 (Mason et al. 1998)
%%Tokovinin \& Ismailov 1988)
and is known to be a bright
X-ray source (Bergh\"{o}fer \& Schmitt 1994; Bergh\"{o}fer, Schmitt,
\& Cassinelli 1996). The A and B components cannot be separated
by existing X-ray telescopes. At a distance of 352 pc, the
angular separation equates to a projected linear separation of
88 AU, or 2101 R$_{O-star}$ (Table 1). 
The extinction toward $\sigma$ Ori AB 
is low with E(B-V) = 0.05 - 0.07
(Table 1), thus minimizing interstellar absorption of any 
soft X-rays that might be produced in stellar wind shocks. 
In terms of wind strength, $\sigma$ Ori A provides a good
comparison with observations of other OB stars. Its wind
is much weaker than the O4f star $\zeta$ Pup, and its
mass-loss rate falls roughly in between that of the moderate
wind source $\delta$ Ori and the weak wind source $\beta$ Cru.

There is additional interest in $\sigma$ Ori because it is
surrounded by a cluster of several hundred young stars,
originally discovered in {\em ROSAT} X-ray observations
(Walter et al. 1997). 
{\em Spitzer} mid-IR observations have identified more than 300
cluster members, with a disk fraction of $\approx$35\% in 
low-mass stars  (Hern\'{a}ndez et al. 2007). A similar
disk fraction was found by Oliveira et al. (2006) from
ground-based K and L$'$-band images. A recent analysis of
{\em TYCHO} photometry by  Mayne \& Naylor (2008) gives
a nominal cluster age of $\sim$3 Myr and a distance
of 389$^{+34}_{-24}$ pc. Some stars are still in the 
accretion phase but the estimated accretion rates are
low, with typical values $\dot{M}$$_{acc}$ $<$ 
10$^{-8}$ M$_{\odot}$ yr$^{-1}$ (Gatti et al. 2008).
This may have some effect on  X-ray emission, which is   
of primary interest here, given that several studies
suggest that the X-ray luminosities of accreting classical
T Tauri stars are on average about a factor of $\sim$2 
lower than their non-accreting counterparts, the weak-lined
T Tauri stars (Preibisch et al. 2005; Franciosini, 
Pallavicini, \& Sanz-Forcada 2006, hereafter FPS06; 
Telleschi et al. 2007).

X-ray observations of the $\sigma$ Ori region  have 
also been obtained with  {\em XMM-Newton}
and 174 X-ray sources were  detected, 76 of which were identified 
with  cluster members
(Sanz-Forcada, Franciosini,\& Pallavicini 2004;  FPS06).
A 98 ksec {\em Chandra} HRC-I
observation (ObsId 2560) has also been acquired, 
detecting more than 140 X-ray sources (Wolk et al. 2006) 
including faint X-ray emission from the infrared source IRS 1
located just  3.$''$3 north of $\sigma$ Ori AB 
(van Loon \& Oliveira 2003; Caballero 2007b).

We present  results of a sensitive {\em Chandra} X-ray
observation of the $\sigma$ Ori cluster obtained
with the High Energy Transmission Grating (HETG)
and the ACIS-S detector. This observation provides
higher spatial  resolution than the previous 
{\em XMM-Newton} observation, but over a smaller
field-of-view. The spatial resolution acquired here
is comparable to that of the   {\em Chandra}
HRC-I observation, but the HETG/ACIS-S combination
provides spectral information whereas HRC-I does not.
On the other hand, the HRC-I observation is  more sensitive,
covers a larger field-of-view, and gives superior image
quality. Thus, the existing
{\em Chandra} and  {\em XMM-Newton} data sets are complementary.

Our primary objective was to obtain a high-resolution
X-ray spectrum of $\sigma$ Ori AB to test wind shock
models of X-ray emission in massive stars. We focus
on spectral properties of $\sigma$ Ori AB but also
analyze zeroth order CCD images, spectra, and light
curves of 41 other X-ray sources near $\sigma$ Ori AB,
including the magnetic helium-strong B2Vpe star
$\sigma$ Ori E (HD 37479) and the unusual infrared source
IRS1.

\section{Chandra Observations and Data Reduction}

The {\em Chandra} observation (ObsId 3738) began on 2003
August 12 at 20:59:02 UT and ended on August 13 at 23:29:25 UT,
yielding 90,967 s of usable exposure.  Pointing was
centered near  $\sigma$ Ori AB (HD 37468) at  nominal
pointing coordinates (J2000.0) 
R.A. = 05$^{\rm h}$38$^{\rm m}$44.$^{\rm s}$43,
decl. = $-$02$^{\circ}$35$'$25.$''$9. Grating spectra 
of $\sigma$ Ori AB were
obtained with the High-Energy Transmission Grating (HETG)
using the  ACIS-S detector in faint timed event mode
with 3.2 s frame times.

The Level 1 events file provided by the {\em Chandra} X-ray
Center (CXC) was  processed using CIAO v. 3.0 
\footnote{Further information on {\em Chandra} Interactive
Analysis of Observations (CIAO) software can be found at
http://asc.harvard.edu/ciao.} using standard science 
threads. The  CIAO processing applied calibration updates
(CALDB v. 2.26), selected good event patterns, removed streak 
artifacts from CCD-8,
determined zero-order source centroids,  defined
grating arm masks and extracted grating spectra,
and created response matrix files (RMFs) and auxiliary
response files (ARFs) for the extracted spectra.
The spectral extraction provides separate dispersed spectra for the
medium-energy grating (MEG) and high-energy grating (HEG).
The HEG provides higher spectral
resolution ($\Delta$$\lambda$ = 12 m\AA~ FWHM, 1.2 - 15 \AA~) than the
MEG ($\Delta$$\lambda$ = 23 m\AA~, 2.5 - 31 \AA~) but the MEG
gives a higher signal-to-noise  ratio (S/N) in 
1st order spectra for $\sigma$ Ori AB. We focus here on
the higher S/N data obtained in MEG 1st order spectra,
combining the $+1$ and $-1$ orders for spectral analysis.

Additionally, we used the CIAO {\em wavdetect} tool to
identify X-ray sources on the  ACIS-S3 chip
(CCD-7) which covers a $\approx$8.3$'$ $\times$ 8.3$'$ region 
near the primary target $\sigma$ Ori AB.  
We ran {\em wavdetect} on full-resolution 
(0.$''$49 pixels) images using events in the 0.5 - 7 keV
range to reduce the background. The {\em wavdetect}
threshold was set at $sigthresh$ = 1.5 $\times$ 10$^{-5}$ 
and scale sizes of 1,2,4,8, and 16 were used.
We visually compared the   {\em wavdetect} source list
against the ACIS-S image and the archived 
HRC-I image to check for spurious and missed detections.

CIAO {\em psextract}
was used to extract zero-order source and background spectra 
for brighter sources ($>$100 net counts).
The nominal  extraction radius used was 
R$_{e}$ = 4 pixels (= 1.$''$97). Zero order light curves were also
extracted for brighter sources using CIAO tools.
Spectral and timing analysis were undertaken
with the HEASOFT {\em Xanadu} 
\footnote{http://heasarc.gsfc.nasa.gov/docs/xanadu/xanadu.html.}
software package including XSPEC v. 12.3.0. We also used
a modified version of the ISIS v. 1.4.8
\footnote{http://space.mit.edu/csc/isis/}
spectral analysis package to complement the XSPEC analysis.
The ISIS implementation allows the flux ratios of 
the resonance (r), intercombination (i), and forbidden (f)
lines in He-like ions to be varied as free parameters 
when fitting grating spectra. This is an advantage
for OB stars since their f/i line flux ratios can be affected by the 
strong UV radiation field. The modified ISIS package
is discussed in more detail by 
Cohen et al. (2008).

\section{X-ray Source Identification and Properties}

\subsection{Source Identification}

Our analysis of zero-order data focuses on sources
detected on the ACIS-S3 chip (CCD7) which contains 
the primary target $\sigma$ Ori AB.  Sources detected
on the other chips in the ACIS-S array lie 
far off-axis and are heavily vignetted and will
not be discussed here. Figure 1
shows the positions of the 42 sources detected on 
ACIS-S3 and their properties are summarized in Table 2.
All of the sources in Table 2 are also visible in the
archived HRC-I image (obsId 2560).

Figure 2 is a zoomed view of the ACIS-S3 image 
in the vicinity of  $\sigma$ Ori AB,  the
brightest X-ray source on ACIS-S3.
The infrared source IRS1 lying 3.$''$3
north of $\sigma$Ori AB is also detected (CXO 20).
Other noteworthy detections are the 
magnetic B2Vpe star $\sigma$ Ori E (CXO 23; Sec. 3.7.2)
and the B5V star HD 37525 (CXO 42), the latter being
a known binary (Sec. 3.7.3). Interestingly,
we do not detect the B2V star $\sigma$ Ori D
even though its spectral type is the same as
$\sigma$ Ori E, a clue that $\sigma$ Ori E
is anomalously bright in X-rays for its spectral 
type. We do note however that 
an X-ray source at or near the position 
of $\sigma$Ori D is visible in the more
sensitive HRC-I image. The A-type star
$\sigma$ Ori C was not detected.

All but five of the X-ray sources in Table 2 have optical or
near-IR counterparts as determined from a search of
the HEASARC object catalogs
\footnote{http://heasarc.gsfc.nasa.gov/docs/archive.html}
using a 2$''$ search radius.
As Table 2 shows, the offsets between the X-ray and
IR/optical positions are in all cases $<$1$''$  
and in  most cases $<$0.$''$5. Interestingly, no
IR counterpart was found for CXO 8, which is one
of the brighter X-ray sources.  However, a faint 
V = 20.88 mag optical source was found in
Mayne et al. (2007) at an offset of 0.$''$18 from
the X-ray position. The X-ray source has an unusually hard
spectrum (discussed below) and is variable.
Further deep  optical and infrared observations would 
be useful to elucidate its nature.

\subsection{Faint Detections and Detection Limit}

The faintest ACIS-S X-ray detection (CXO 32) has
5 $\pm$ 2 counts and is  confirmed  in the HRC-I image.
Assuming a  slightly more conservative 7 count
detection threshold,  the corresponding X-ray
luminosity threshold determined from PIMMS
\footnote{For information on PIMMS (Portable Interactive
 Multi-Mission Simulator) see 
http://asc.harvard.edu/ciao/ahelp/pimms.html.}
at a cluster distance d $\approx$ 390 pc is 
log L$_{x}$ = 28.81 ergs s$^{-1}$ (0.5 - 7 keV). Here
we have assumed a typical late-type pre-main-sequence star
modeled as an isothermal X-ray source with  a 
Raymond-Smith plasma at kT = 2 keV and an absorption
column density 
N$_{\rm H}$ = 3 $\times$ 10$^{20}$ cm$^{-2}$
(E(B-V) $\approx$ 0.05; Gorenstein 1975 conversion).

At a luminosity threshold log L$_{x}$ = 28.81 ergs s$^{-1}$ 
we would expect to detect X-ray emission from pre-main
sequence stars in Orion down to masses of  $\sim$0.2 - 0.3 M$_{\odot}$,
based on previous deep {\em Chandra} Orion observations 
(Figs. 3 and  7 of Preibisch et al. 2005). A comparison of
the X-ray detections in Table 2 with the optical catalog
of Sherry, Walter, \& Wolk (2004, = S04) confirms 
this. Two low-mass detections are S04 15 = CXO 17 and
S04 18 = CXO 31, both of which have high cluster membership
probability (92\% - 93\%). Their mass estimates
from S04 are 0.22 M$_{\odot}$ (S04 15) and 
0.18  M$_{\odot}$ (S04 18) and their respective
V - R$_{c}$ colors imply equivalent main-sequence
spectral types of $\sim$M4V -  M5V (Kenyon \& Hartmann 1995).
Also noteworthy is {\em Chandra} source CXO 22, which was
classified as M5 by FPS06 but was listed in their Table A.2
(object name B3.01-67) as undetected by {\em XMM-Newton}.

Some of the faint sources in Table 2 are likely extragalactic
background objects. In particular, the five sources without 
optical or near-IR identifications are extragalactic candidates
since all are faint  ($\leq$25 counts), non-variable, 
and have above-average mean photon energies. If we assume 
a typical extragalactic source
power-law X-ray spectrum with a photon power-law index $\Gamma$ = 1.4,
N$_{\rm H}$ = 3 $\times$ 10$^{20}$ cm$^{-2}$, 
and a 7-count detection threshold, then the hard-band (2 - 8 keV)
number counts from the {\em Chandra} Deep Field North (DFN)  
observations (Cowie et al. 2002) predict $\approx$15 extragalactic 
sources in the ACIS-S3 CCD field-of-view above our detection limit. 
The expression for hard-band (2 - 8 keV) number counts obtained from 
DFN data by Brandt et al. (2001) gives a similar result.
However, the accuracy
of the log $N$ - log $S$ distribution for extragalactic sources
from the Deep Field observations toward lower galactic latitude
star-forming regions such as $\sigma$ Ori ($b$ $\approx$ $-$17$^{\circ}$)
is not well-known. For the deep {\em Chandra} COUP observation of
the Orion Nebula Cluster ($b$ $\approx$ $-$19$^{\circ}$),
Getman et al. (2005) concluded that the number of extragalactic
sources in the X-ray sample was a factor of $\sim$3 - 4 less
than predicted by {\em Chandra} Deep Field number counts.

\subsection{Hardness}

The typical
mean photon energy (Table 2) of detected sources is
$<$E$>$ = 1.83 keV and the distribution of $<$E$>$ values
is shown in Figure 3.  The brightest X-ray source
$\sigma$ Ori AB   also has the lowest
mean energy  $<$E$>$ = 0.86 keV. The hardest emission is from
the  very faint source CXO 32 with $<$E$>$ = 4.06 keV, for
whicn no optical or IR counterpart was found. These properties 
suggest that it may be an extragalactic background source.

\subsection{Variability}

X-ray variability is frequently detected  in T Tauri stars.
The most extensive data on X-ray variability for young stars in
Orion come from the {\em Chandra} Orion Ultradeep Project (COUP), 
which obtained  a long $\sim$13 day  observation of the Orion 
Nebula Cluster (ONC). Rapid X-ray  variability ($\sim$minutes
to hours) associated with  magnetic-reconnection flares was
detected in T Tauri stars (Favata et al. 2005), as well as
slower X-ray modulation linked to stellar rotation at periods of
$\sim$2 - 12 days (Flaccomio et al. 2005). A variability  analysis 
of 20 massive O, B, and A stars detected in the COUP observation
was undertaken by Stelzer et al. (2005). Significant variability 
was detected in at least 6 of the 9  earliest O7 - B3 stars having  
strong winds. These include known multiple systems which appear to 
show magnetic behavior such as  $\theta^1$ Ori A (B0), 
$\theta^2$ Ori A (O9.5), and $\theta^1$ Ori C (O7). But, no
variability was seen in stars showing the soft X-ray spectra
typcially associated with radiative wind shocks such as 
$\theta^1$ Ori D (B0.5) and NU Ori (B1). {\em Chandra} observations 
of other massive stars in Orion have also revealed objects with
soft X-ray spectra and no detectable variability, such as
$\delta$ Ori (Miller et al. 2002).

Nine sources detected in our $\sigma$ Ori observation
have high probability of variability as gauged by
P$_{var}$ $\geq$ 0.95 based
on the Kolmogorov-Smirnov (KS) test (Press et al. 1992) 
using unbinned photon arrival times. These variable sources
are identified in column 4 of Table 2. The X-ray light curve
of $\sigma$ Ori AB (Fig. 4) shows no evidence
of large-amplitude variability. A KS test
on unbinned events gives  a variability probability 
P$_{var}$ = 0.935,
slightly below the limit   P$_{var}$  $\geq$ 0.95
that we adopt for variability.
A $\chi^2$ test on the light curve binned at
4000 s intervals ($\approx$108 counts/bin) gives
P$_{var}$ $<$ 0.001. Thus, we find no
evidence for significant ($\geq$95\% confidence)
X-ray variability in $\sigma$ Ori AB.

The magnetic B2Vpe star $\sigma$Ori E is variable (Fig. 5;
see also Sec. 3.7.2). However, no large flares were
detected, in contrast to the large outbursts seen
in the {\em XMM-Newton} observation (Sanz-Forcada et al. 2004)
and in the {\em Chandra} HRC-I exposure.
Slow low-amplitude
X-ray variability is seen in the unusual 
hard source CXO 8 (Fig. 6).  
X-ray flares are clearly visible in CXO 9 (Fig. 7) 
and CXO 25 (Fig. 8). The counterpart of CXO 9
has been classified as  K8 and has properties of
a weak-lined TTS (Zapatero Osorio et al. 2002). CXO 25 is
associated with a M4 star that has a near-IR
excess (Oliveira et al. 2006) indicative of
a residual disk or envelope. The second brightest
X-ray source in Table 2 (CXO 37) showed a slow
decline in count rate by a factor of two during
the observation, but no large flares.

\subsection{Near-IR Colors}

Figure 9 shows J-K$_{s}$ colors of those 
X-ray detections with 2MASS identifications.
IRS1 was not resolved from $\sigma$ Ori AB
by 2MASS and no published JHK magnitudes are available.
Most of the sources lie in the range
0.7 $\leq$ J-K$_{s}$ $\leq$ 1.3, typical of
early K to late M dwarfs or giants 
(Bessell \& Brett 1988). The three massive
stars $\sigma$ Ori AB, $\sigma$ Ori E, and 
the B5V star HD 37525  are
clearly separated in color-magnitude space
by their smaller J-K$_{s}$ and K$_{s}$ values.
An interesting outlier is the faint hard variable 
X-ray source CXO 12 which has
a large value J-K$_{s}$ = 2.33 and is the faintest
near-IR identification with  K$_{s}$ = 14.04.
It is associated with a faint V = 18.8 mag optical
source (Mayne et al. 2007).

Figure 10 is a color-color diagram based on 
2MASS JHK$_{s}$ photometry of X-ray detections
with 2MASS identifications. Excluding the three
known high-mass stars, nineteen of the X-ray sources 
in Figure 10 have assigned spectral types (FPS06)
and all of these are K or M stars. This substantiates
the above conclusion that most of the low-mass X-ray detections
are  K or M stars. Of the nineteen X-ray sources with
known spectral types, slightly more than half lie
to the right of their respective reddening lines as
determined from the intrinsic colors of
Bessell \& Brett (1988) and the extinction law of
Rieke \& Lebofsky (1985). Thus, in this limited
sample of 19 stars, roughly half show evidence of
near-IR excesses based on existing spectral
type assignments and {\em 2MASS} colors.

\subsection{CCD Spectra}

We have fitted the zero-order CCD spectra of brighter sources 
($\geq$100 counts) with  one-temperature (1T) or two-temperature (2T)
{\em apec} solar-abundance optically thin plasma models 
(Smith et al. 2001) in XSPEC, as summarized
in Table 3.  We have also fitted the spectrum of the high-interest
object  IRS1 (CXO 20), but the fit is not well-constrained because 
this source is faint (43 counts). 
Photon pileup is low ($\approx$5\%) in the zero-order 
spectrum of $\sigma$ Ori AB (neglibigle in first order) and
zero-order pileup is  negligible ($<$1.7\%) in all the other
sources.

In all but a few sources there were insufficient counts to constrain both 
the temperature and absorption column density, so we held the 
absorption fixed at N$_{\rm H}$ = 3 $\times$ 10$^{20}$ cm$^{-2}$,
consistent with  E(B$-$V) $\approx$ 0.05 mag for  the 
$\sigma$ Ori region. The best-fit value of  N$_{\rm H}$ was
within a factor of $\approx$2 of the above value for the brighter
sources where N$_{\rm H}$ was allowed to vary during the fit.

Two-temperature models provide an acceptable approximation
to the spectra in most cases. With the notable exception
of $\sigma$ Ori AB (discussed below), the derived temperatures from 2T models
are typical of magnetically-active pre-main sequence stars
in the Orion region (Preibisch \& Feigelson 2005), showing 
both a cool component (kT$_{1}$ $<$ 1 keV) and  
hotter plasma that is usually in the range 
kT$_{2}$ $\approx$ 2 - 3 keV.

%%We held the metal abundances fixed at solar values
%%in fitting the zero-order spectra due to limited counts.
%%If the global metal abundance is allowed to vary in
%%the $\sigma$ Ori AB fit then the fit is improved
%%and converges to Z = 0.25 [0.15 - 0.38; 90\% conf.]
%%Z$_{\odot}$ (Table 3). More detailed information on abundances
%%is given below as derived from higher resolution
%%grating spectra. 

\subsection{Comments on Specific Stars}

\subsubsection{HD 37468 ($\sigma$ Ori AB)}

Fits of the CCD spectrum of $\sigma$ Ori AB 
excluded events with energies below 0.5 keV to avoid
possible contamination from optical light leaks
\footnote{The ACIS CCDs are sensitive at optical frequencies
but most optical light is diffracted away from the CCDs
by the optical blocking filter. Also, the optical sensitivity
is further reduced when the MEG/HEG gratings are in place.
Further details can be found at
http://www.astro.psu.edu/xray/docs/cal\_report/node188.html.}.
A simple 1T $apec$ optically thin plasma model (1T $apec$)
was not acceptable.
But, a 2T $apec$ model does give an acceptable fit (Fig. 11) 
with two cool components   at  
kT$_{1}$ = 0.21 keV
and kT$_{2}$ = 0.47 keV whose emission-measure weighted average
is kT = 0.3 keV. A fit with the differential emission measure 
model $c6pmekl$ in XSPEC using solar abundances gives
good results ($\chi^2$/dof = 43.4/44) and  shows a single broad 
peak in the emission measure  between 0.2 - 0.3 keV 
with significant emission measure over the 0.15 - 0.45 keV 
range (Fig. 12). Thus, the  2T model is probably an approximation to
this broad cool emission measure distribution.
The spectrum of  $\sigma$ Ori AB is clearly dominated by cool
plasma, consistent with expectations for radiative wind shocks.

As noted above, {\em Chandra} does not spatially resolve the 
A and B components. Thus, our observation does not exclude
the possibility that the X-ray emission is the superposition of
soft emission from the  O9.5V primary and the B0.5V secondary.
Soft X-ray emission from some B-type stars in Orion has previously 
been reported, including the B0.5 star $\theta^1$ Ori D (Stelzer
et al. 2005).

\subsubsection{HD 37479 ($\sigma$ Ori E)}

The helium-strong star $\sigma$ Ori E has been extensively
studied and is known to exhibit optical, UV, and radio
modulations at a period P = 1.1908 days (Reiners et al.
2000), which has been interpreted as the stellar rotation period.
A strong magnetic field that varies from $-$2.2 to $+$2.8  kG 
at the  1.19 day period was detected by 
Landstreet \& Borra (1978) and discussed further by
Borra \& Landstreet (1979). 
A distance of 640 pc was obtained by
Hunger, Heber, \& Groote (1989) which, if correct,
would place it behind the $\sigma$ Ori cluster.

An X-ray flare detected by  {\em ROSAT} was reported
by Groote \& Schmitt (2004), who concluded that the 
flare likely occurred on the magnetic B star and not
on a hypothesized late-type companion. A large X-ray
flare whose rise and decay spanned at least $\approx$8 
hours was also detected by {\em XMM} (Sanz-Forcada et al. 2004).
The X-ray flare CCD spectrum recorded by the PN camera
showed fluorescent Fe emission at $\approx$6.4 keV,
a signature of cool neutral or near-neutral material
near the star that was irradiated by hard X-rays 
during the flare. Sanz-Forcada et al. argued that
an unseen low-mass companion was likely responsible
for the flare. A large flare is also visible in the 
archived {\em Chandra} HRC-I data.

The {\em Chandra} ACIS zero order light curve (Fig. 5) is clearly 
variable but no large-amplitude flares were detected.
A KS test gives a variability probability  
P$_{var}$ = 0.986. Low-level modulation may
be present in the ACIS light curve, as shown in Figure 5.
However, the {\em Chandra} observation spanned only 
88\% of the 1.19 day rotation period and additional
time monitoring over more than one period would be
needed to substantiate any periodic X-ray modulation.

The {\em Chandra} ACIS spectrum of $\sigma$ Ori E 
(Fig. 13) is definitely harder than that of
$\sigma$ Ori AB, as anticipated from  their
respective mean photon energies (Table 2).
However, the ACIS spectrum is not as hard as
seen by {\em XMM} during the flare state and we 
do not detect the fluorescent Fe line at
6.4 keV. A 2T $apec$ model of the ACIS spectrum
gives a marginally acceptable fit with a 
high-temperature component at kT$_{2}$ = 2.4 keV,
which is typical of magnetically-active stars.
The fit can be  improved  by using a 3T model
and allowing N$_{\rm H}$ to vary (Table 3 Notes),
and this fit is overlaid in Figure 13.
The 3T model gives a temperature kT$_{3}$ = 
2.6 [2.1 - 3.3; 90\% confidence] keV
for the hot component, and the inferred value of
N$_{\rm H}$ is consistent with E(B$-$V) = 0.05 (or possibly larger)
estimated for this star by Hunger et al. (1989).

The stellar luminosity from the blackbody relation
is  L$_{*}$  = 
2.48 $\times$ 10$^{37}$ ergs s$^{-1}$,
or log (L$_{*}$/L$_{\odot}$) = 3.81,
where we have assumed T$_{eff}$ = 22,500 K and
R$_{*}$ = 5.3$^{+1.35}_{-1.1}$ R$_{\odot}$ from  Hunger et al. (1989).
The unabsorbed flux from the 3T model gives
an X-ray luminosity 
L$_{x}$ (0.5 - 7 keV) = 2.88 $\times$ 10$^{25}$d$_{pc}^2$,
where d$_{pc}$ is the distance in parsecs.
Adopting d = 640 pc (Hunger et al. 1989)  gives 
log (L$_{x}$/L$_{*}$) = $-$6.32. This ratio is
much higher than usually found for B2V stars, for
which a catalog of {\em ROSAT} detections
gives typical values  log (L$_{x}$/L$_{*}$) 
$\approx$ $-$7.2 to $-$6.8 (Bergh\"{o}fer et al. 1996). 
In fact, these typical values are too large since the
{\em ROSAT} All Sky Survey (RASS) was based on
short exposures and thus selected in favor of the detection of
X-ray bright B2V stars. A ratio  log (L$_{x}$/L$_{*}$)
$\approx$ $-$8 may actually be more representative
(Cohen, Cassinelli, \& MacFarlane 1997).
This strengthens the conclusion
that the value of L$_{x}$/L$_{*}$  for $\sigma$ 
Ori E at d = 640 pc is anomalously
large for a B2V star. This is a clue that the 
star may actually lie in the $\sigma$ Ori cluster
instead of behind it.

If $\sigma$ Ori E lies in the cluster at d $\sim$ 390 pc,
then log (L$_{x}$/L$_{*}$) = $-$6.75. This is in better
agreement with other B2V stars, but still at the high
end of the range.  Taking into account the uncertainty in 
the X-ray  flux  measurement from the 3T $apec$ fit ($\approx$15\%),
the uncertainty in R$_{*}$ (see above), and allowing for a slightly 
higher temperature  T$_{eff}$ = 23,500 adopted in some studies
(Nakajima 1985; Groote \& Schmitt 2004), the  above ratio
could be as  low as log (L$_{x}$/L$_{*}$) = $-$7.09.
Thus, uncertainties could conspire to decrease the 
L$_{x}$/L$_{*}$ ratio to a value that is within the 
range found for X-ray bright {\em ROSAT} B2V star detections.

In summary,  the L$_{x}$/L$_{*}$ ratio for 
$\sigma$ Ori E is anomalously high compared to other B2V 
stars if it lies at d = 640 pc. If the star lies in the cluster 
at d = 390 pc, the ratio is in better agreement with that of
other X-ray luminous B2V stars, but still at the high end of
the range if the nominal stellar parameters are adopted.
These indications that the  L$_{x}$/L$_{*}$ ratio is high
and persistent X-ray flaring leave open the possibility that
a late-type companion may be contributing to the X-ray emission
of $\sigma$ Ori E. However, there have been no reports of
the detection of a companion so far.

The rather high plasma temperature for this
star and its X-ray variability are not consistent
with radiative wind shock predictions. The
detection of a strong magnetic field gives reason
to believe that the magnetically-confined wind shock (MCWS) 
picture may be relevant. This model is capable of
producing higher plasma temperatures than  
the conventional radiative  wind shock scenario
and was originally formulated to explain hot
plasma in stars with strong $\sim$kG magnetic
fields such as the Ap star IQ Aur
(Babel \& Montmerle 1997a = BM97a). It has also
been invoked to explain hard rotationally-modulated
X-ray emission in the young magnetic O star
$\theta^1$ Ori C (Babel \& Montmerle 1997b; 
Donati et al. 2002; Gagn\'{e} 
et al. 2005; Schulz et al. 2000). The good agreement
between model predictions and observed X-ray parameters
for $\theta^1$ Ori C found by Gagn\'{e} et al. (2005) 
provides solid support for the MCWS interpretation. 

In the MCWS picture, a strong surface magnetic field
can channel the wind flow into two oppositely
directed streams from each hemisphere, which collide
near the  magnetic equator and liberate their 
kinetic energy in shock-heated plasma.
The degree to which the wind is confined by the
magnetic field is expressed in terms of the 
confinement parameter 
$\eta$ = B$_{eq}^2$R$_{*}^2$/$\dot{M}$v$_{\infty}$
where B$_{eq}$ is the equatorial  magnetic field 
strength  (ud-Doula \& Owocki 2002).
For B $\sim$ 2 kG and plausible mass-loss
parameters (Groote \& Hunger 1997), one
obtains  $\eta$ $>>$ 1 for $\sigma$ Ori E.
That is, the wind is strongly confined by
the magnetic field.

Building on earlier work, Townsend, Owocki,
\& Groote   (2005)
extended the MCWS idea into a rigidly rotating
magnetosphere  model. They showed that 
by allowing  the center of a dipole magnetic
field to be displaced away from the center of
the star, many aspects of the periodic behavior
of $\sigma$ Ori E could be explained. This
work has now been expanded into a rigid field
hydrodynamics (RFHD) approach by Townsend, Owocki,
\& ud-Doula (2007). The RFHD  approach is relevant to
stars with strong magnetic fields such as $\sigma$ Ori E 
where  the wind has little effect on the  field lines. 

Numerical RFHD simulations of the X-ray emission of
a generic star with properties similar to $\sigma$ Ori E 
by Townsend et al. (2007) predict hot plasma
with an emission measure distribution peaking
near kT $\approx$ 3.5 keV. This is somewhat
hotter than we infer from  our best-fit 3T
model, which gives kT$_{3}$ = 2.6 [2.1 - 3.3; 90\% conf.]
keV. However, as Townsend et al. (2007) have noted,
the higher temperature in the RFHD model could be
a consequence of  uncertainties in the mass loss
parameters and neglecting thermal conduction effects.
Future  refinements of the RFHD
model tailored specifically to $\sigma$ Ori E
will address these temperature differences
and also have the potential to provide
theoretical estimates of the X-ray luminosity
and X-ray absorption that can be directly compared
with observations.

\subsubsection{HD 37525}

This star is classified as B5V and has a known 
companion at a separation of 0.$''$45 $\pm$ 0.$''$04
(Caballero 2005). The {\em Chandra} ACIS-S spatial resolution
is not sufficient to distinguish between the primary
and the companion. But, if the companion is a pre-main
sequence star (Caballero 2005), then it likely contributes
to the X-ray emission.

\subsubsection{$\sigma$ Ori IRS1}

IRS1 lies 3.$''$3 north-northeast of $\sigma$ Ori AB
and was discovered in sub-arcsecond mid-IR observations
with the ESO 3.6 m telescope by
van Loon \& Oliveira (2003). It shows a clear near to
mid-IR excess indicative of a disk. The mid-IR emission
was  resolved into an elongated structure extending away
from  $\sigma$ Ori AB that was described as 
``fan-shaped'' by van Loon \& Oliveira (2003).
The unusual elongated structure and small angular separation
between IRS1 and $\sigma$ Ori AB support the conclusion
that the two objects are physically associated and that
circumstellar material around IRS1 is being 
influenced by the strong OB system radiation field.

Observational evidence for the source now known as IRS1
was in fact  already present in earlier {\em VLA} radio 
observations obtained by  Drake (1990). No radio source was 
detected at the optical position of $\sigma$ Ori AB, but he
reported a radio source at J2000 position
RA = 5$^{\rm h}$ 38$^{\rm m}$ 44.78$^{\rm s}$,
decl. = $-$2$^{\circ}$ 35$'$ 57.$''$6. This  
source lies $\approx$2.$''$7 north-northeast of  
$\sigma$ Ori AB, but is offset by only $\approx$1$''$ from 
the IRS1 8.6 $\mu$m peak (Fig. 1 of van Loon \& Oliveira 2003).
Thus, an association between the radio source and
IRS1 seems likely.

The 6 cm flux densities reported by Drake (1990) are
much larger than would be expected for free-free wind
emission from $\sigma$ Ori AB, and this provides 
additional support that the radio source is not 
$\sigma$ Ori AB.
Adopting the wind parameters for $\sigma$ Ori AB
in Table 1, the predicted 6 cm radio flux density 
for  free-free emission from a spherical fully-ionized
wind is S$_{4.86 GHz}$ = 0.1 mJy (eq. [7] of
Skinner, Brown, \& Stewart 1993). This value is a 
factor of 15 - 20 smaller than the observed flux
density of the detected radio source (Table II of  
Drake 1990). The upper limit on the radio flux density
measured at the position of $\sigma$ Ori AB in {\em VLA} 
images is   S$_{4.86 GHz}$ $\leq$ 0.3 mJy (S. Drake,
private communication), which  is consistent
with the predicted 0.1 mJy flux density for  free-free 
wind emission.

{\em Chandra's} high angular resolution clearly reveals
a faint X-ray source (CXO 20) located 2.$''$85 north
and 0.$''$93 east of $\sigma$ Ori AB, which we associate
with IRS1 (Fig. 2). Only 43 net counts were detected from 
this X-ray source, 
but this provides enough information to undertake a rudimentary
spectral analysis. The CCD spectrum (Fig. 14) is  absorbed
below $\approx$0.7 keV and peaks at E $\approx$ 1 keV.
Spectral fits imply hotter plasma at kT $>$ 1 keV (Table 3).
The unabsorbed X-ray  luminosity of IRS1 is 
L$_{x}$ (0.5 - 7 keV) = 2.75 $\times$ 
10$^{24}$ d$_{pc}^2$ (ergs s$^{-1}$),
where d$_{pc}$ is the distance in parsecs. Assuming 
IRS1 is physically associated with $\sigma$ Ori AB at
a distance of $\sim$352 pc, one obtains 
log L$_{x}$ (0.5 - 7 keV) = 29.53 ergs s$^{-1}$.
The KS test gives a high variability probability 
P$_{var}$ = 0.95 for IRS1. A coarse light curve 
binned at 11000 s intervals (9 bins with $\approx$5 counts/bin)
shows a slow increase in count rate by a factor of $\sim$3
during the last 70 ksec of the observation, but with large
count-rate uncertainties because of the faint emission.  
The likely presence of  variable X-ray emission from IRS1 at
a plasma temperature of kT $\approx$ 1 keV (and perhaps
higher) demonstrates  that the infrared source detected by
van Loon \& Oliveira harbors a magnetically-active young  star.

The {\em Chandra} COUP study has shown that the X-ray luminosity
of optically-revealed pre-main sequence  (PMS) stars in the Orion 
nebula correlates with mass, albeit with large scatter 
(Preibisch et al. 2005). Assuming a similar correlation
for the $\sigma$ Ori cluster, the above   L$_{x}$
along with the mid-IR data would imply that IRS1 is 
likely a T Tauri star with a mass of a few tenths of 
a solar mass whose disk/envelope is being photo-evaporated 
in the harsh UV environment of $\sigma$ Ori AB.

\subsubsection{Source CXO 8 (J053835.21$-$023438.1)}

A thermal plasma model requires a very high temperature
for the variable source CXO 8, but a simple absorbed power-law model
also gives an acceptable   fit (Table 3). We find no counterpart in
HEASARC extragalactic catalogs but the X-ray source lies at a small
offset from a faint V = 20.88 mag optical source  (Mayne et al. 2007).

\section{Grating Spectrum of $\sigma$ Ori AB}

\subsection{General Properties}
Figure 15 shows the first-order MEG1 spectrum of
$\sigma$ Ori AB with prominent lines identified.
Line properties are summarized in Table 4. 

In order to test the feasibility of various emission
models (Sec. 5.), we need information on X-ray
temperature, line widths, line centroid shifts
(if any), and flux ratios of specific lines. 
Information on line shape is also important,
since radiative wind shock models predict 
line asymmetries for optically thick winds. 
But at the signal-to-noise ratio
of the MEG1 spectrum, the lines can be 
acceptably  fitted with Gaussians and we  do
not find any conclusive evidence of line asymmetries.

The spectrum is clearly dominated by lines
from relatively cool plasma, with 
the coolest lines being the O VII He-like
triplet $r$ and $i$ lines and the 
O VII line at 18.627 \AA~ 
(log T$_{max}$ = 6.3 K). The highest
temperature and shortest wavelength line detected was 
Si XIII ($\lambda_{lab}$ = 6.6479 \AA;
log T$_{max}$ = 7.0 K).

\subsection{Line Widths}
Line fluxes and widths were based on fits with a 
weak power-law component plus a Gaussian profile convolved 
with the instrumental response. The weak power-law
component accounts for faint emission outside
the line, due either to true continuum or 
a pseudo-continuum arising from many faint
unresolved lines. 

The strongest lines
based on observed line flux are O VIII 
($\lambda_{lab}$ = 18.967 \AA) and the O VII
intercombination line ($\lambda_{lab}$ = 21.804 \AA).
In general, all lines for which widths could be
reliably measured are moderately broadened in
excess of thermal (Doppler) widths, the latter being 
no greater than $\sim$60 km s$^{-1}$ for
the detected ions at a temperature T $\sim$ 3.5 MK.
 
Gaussian fits of nine prominent  lines
give a mean  half-width  at half-maximum
$\overline{\rm HWHM}$ =  264$\pm$80 (1$\sigma$) km s$^{-1}$. 
Spectral fits with ISIS give nearly identical results.
Taking the uncertainties
of the line width measurements into  account, there is no 
clear dependence of line width on wavelength or 
maximum line power temperatures T$_{max}$
(Figs. 16 - 17).

The mean HWHM for $\sigma$ Ori AB is much less than observed
for O supergiants (Figure 6 of Cohen et al. 2003),
slightly less than for the O9.5 II star $\delta$ Ori 
(HWHM = 430 km s$^{-1}$; Miller et al. 2002),
very similar to that of the B0.2V star $\tau$ Sco
(Cohen et al. 2003), but slightly larger than found
for the B0.5 III star $\beta$ Cru (HWHM $\approx$ 150 km s$^{-1}$;
Cohen et al. 2008). Although there is considerable scatter
in the line widths for any particular star and the sample is
still small, these results suggest tentatively that there
is a trend for broader lines in O stars with stronger winds
versus B stars with weaker winds.
The  mean HWHM value for $\sigma$ Ori AB is well below 
the terminal wind speed 
v$_{\infty}$ $\approx$  1060 km s$^{-1}$ (Howarth and Prinja 1989).
Expressed as a ratio, the  value from individual line fits
gives  $\overline{\rm HWHM}$/v$_{\infty}$ = 0.22 -  0.25.
These ratios are comparable to the value
0.22 found for  $\delta$ Ori  (Miller et al. 2002).

\subsection{Line Centroids}
Gaussian line centroids were measured for 13 stronger lines. 
In all cases the
centroids lie within 7 m\AA ~of the laboratory values
and for eight lines the offsets are  $\leq$3 m\AA,
as shown in Figure 18.
These offsets are well within the $\pm$11 m\AA ~absolute wavelength
calibration accuracy of MEG1
\footnote{{\em Chandra} Proposer's Observatory Guide:
http://asc.harvard.edu/proposer/POG/ .}.
Both positive and negative offsets were found with
ten lines  blueshifted and three redshifted. 
The mean centroid offset was  
$-$1.6 $\pm$ 3.5 ($\pm$1$\sigma$) m\AA.~
In only two cases does $\lambda_{lab}$ lie outside
the $\pm$1$\sigma$ uncertainty interval of the 
measured centroid. The centroid measurements of
these two lines  are uncertain due to close blends 
(Ne X 12.1321/12.1375 \AA)
and possible  contamination by weak nearby Fe lines
(Ne IX  13.553 \AA).

\subsection{Helium-like Ions}
He-like triplets show strong resonance (r) and 
intercombination (i) lines, but the forbidden (f)
lines are very faint and formally undetected.
The resonance and intercombination lines were detected in
the Mg XI (Fig. 19), Ne IX (Fig. 20), and O VII (Fig. 21) He-like
triplets. The Si XIII resonance line was detected but 
the emission in the triplet is so  faint that no reliable
diagnostic information can be obtained apart from an
estimate of the resonance line flux. The strong i lines
and faint or absent f lines are a clear indication that
the f and i line fluxes are being affected by the strong
UV radiation field of $\sigma$ Ori A (T$_{eff}$ = 
33,000 K; Howarth \& Prinja 1989). 

In general, the 
line flux ratio $R$ = f/i is sensitive to both
electron density $n$ and the UV radiation field
(Gabriel \&  Jordan 1969).  
If $\phi$ is the photoexcitation rate 
from the $^{3}$S$_{1}$  level to the  
$^{3}$P$_{2,1}$  levels, then
$R$ = $R_{0}$/[1 $+$ ($\phi$/$\phi_{c}$) $+$ ($n$/$n_{c}$)], where
$\phi_{c}$ and $n_{c}$ are the critical photoexcitation 
rate and critical density. The quantities $R_{0}$, $\phi_{c}$,
and $n_{c}$ (Table 5) depend only on atomic parameters and the electron
temperature of the source, as defined in
Blumenthal, Drake, \& Tucker 1972, hereafter BDT72).

In hot star winds, densities  $n$ $>$ $n_{c}$ high 
enough to affect $R$ would only occur very close to the
star and it is thus reasonable to assume that any
observed reduction in $R$ relative to $R_{0}$ is the result
of photoexcitation by the hot star UV radiation 
field. Thus, in hot stars the value of $R$ is not a 
reliable density diagnostic but is instead a measure of
the photoexcitation rate in the line-emitting region,
and can be used along with atmospheric models to 
constrain the distance from the star at which the
He-like triplet line is formed.
In cool stars with negligible UV flux, the value
of $R$ can be used to constrain the density.
In the limit of low photoexcitation 
($\phi$ $<<$ $\phi_{c}$) and low
density ($n$ $<<$ $n_{c}$),
~$R$ $\rightarrow$ $R_{0}$.
 
For the He-like ions detected in the MEG spectrum of
$\sigma$ Ori AB, the most stringent limit on the 
line formation radius based on the measured upper
limit of $R$ comes from the Mg XI triplet.
We have used a solar abundance TLUSTY model atmosphere
(Lanz \& Hubeny 2007) with T$_{eff}$ = 32,500 and 
log $g$ = 4.0, along with  the formalism in BDT72
and Leutenegger et al. (2006), taking
an average of the photospheric fluxes of the two
$^3$S - $^3$P$_{1,2}$ transitions (997.5 \AA~ and
1034.3 \AA~ for Mg XI). We have averaged these fluxes
over $\approx$3 \AA~ blueward of these wavelengths to 
account for the Doppler shift in the wind  (Leutenegger et al. 
2006).  We find an expression for the ratio $R$(r) = f/i 
as a function of the distance r from the center of the
star, which takes into account the dilution factor W(r) 
and is plotted  in Figure 22.

For Mg XI our spectral fits give a 90\% confidence upper
limit $R$ = f/i $\leq$ 0.17 (Table 5).
As can be seen in Figure 22, this upper
limit implies a small line formation
radius r $\leq$ 1.8 R$_{*}$, where as usual, r is measured
from the center of the star.  A similar analysis for Ne IX
gives $R$ $\leq$ 0.15 and a formation radius r $\leq$ 4.4 R$_{*}$.
The O VII triplet does not provide a useful constraint
on the line formation radius.
Using the adopted mass-loss parameters in Table 1 and
an estimated wind X-ray opacity $\kappa_{w}$ $\approx$ 80 
cm$^{2}$ g$^{-1}$ near E $\approx$ 1 keV 
(Waldron et al. 1998; Cohen et al. 1996; 
Ignace, Oskinova, \& Foullon 2000), 
the wind is expected to be optically thin at E $\approx$ 1 keV  
down to the stellar surface. In addition, our modeling of
the Fe XVII line (15.014 \AA~) confirms that the wind
optical depth is consistent with zero.
Thus, X-rays formed at the small 
radii given above could escape through the wind.

\subsection{Abundances}

We have simultaneously fitted the zero-order spectrum and
the MEG1 and HEG1 first order spectra of $\sigma$ Ori AB 
to search for significant deviations from solar abundances.
These fits were conducted using an absorbed 2T $vapec$ model
with a  modified version of 
ISIS (Cohen et al. 2008) that allows the R and 
G ratios of He-like triplets
to be fitted as free parameters. We allowed the
abundances of elements with prominent lines in the spectrum
to vary, namely  O, Ne, Mg, Si, and Fe. We found
no significant deviations from solar abundances. The
largest deviation found was for magnesium, which gave
a best-fit abundance Mg = 0.78 [0.62 - 1.08; 90\% conf.]
$\times$ solar.

\subsection{Summary of Grating Spectra Results}

In summary, the key properties determined from the 
spectral analysis of $\sigma$ Ori AB are:
(i) dominant cool plasma (kT $\approx$ 0.3 keV), 
(ii) moderate line broadening 
(HWHM $\approx$ 264 km s$^{-1}$)
that is a factor of $\sim$4 below the terminal wind speed of $\sigma$ Ori,
(iii) no significant mean line centroid shifts relative to laboratory
values, (iv) strong resonance and intercombination lines and  
very faint (formally
undetected) forbidden lines with low f/i flux ratios, 
indicative of strong   UV field 
excitation effects, and (v) Mg XI line formation within a few
radii of the star (r $\ltsimeq$ 1.8 R$_{*}$).

\section{Interpretation of the X-ray Emission from $\sigma$ Ori AB}

\subsection{Radiative Wind Shocks}

The temperature of the cool plasma that 
dominates the  X-ray emission  of $\sigma$ 
Ori AB is  in good agreement with model predictions for 
shocks distributed in a  radiatively-driven stellar wind.
Numerical radiation hydrodynamics simulations
of shocks in radiatively-driven winds give typical shock jumps
$\Delta$V = 500 - 1000 km s$^{-1}$ and post-shock temperatures
in the range $\sim$1 - 10 MK (Owocki, Castor, \& Rybicki 1988). This
range can easily accommodate  $\sigma$ Ori AB, whose emission-measure
weighted temperature from the best-fit 2T $apec$ model 
is T $\sim$ 3.5 MK.

The predicted post-shock X-ray temperature corresponding to a shock jump
velocity $\Delta$V is T$_{x}$ = 1.4 $\times$ 10$^{5}$
($\Delta$V/100 km s$^{-1}$)$^2$~K ~(Krolik \& Raymond 1985).
For $\sigma$ Ori AB we have  T $\sim$
3.5 $\times$ 10$^{6}$ K (kT = 0.3 keV) so the required shock jump is
$\Delta$V $\approx$ 500 km s$^{-1}$, or about half the 
terminal wind speed.

Also, the emission measure (EM) in the wind is sufficient to
account for the observed X-ray emission measure. 
The wind emission measure outside a given radius R$_{o}$ is
EM$_{wind}(r >  R_{o}) =  \int_{R_{o}}^{\infty}n_{e}n_{H} 4 \pi r^2 dr$,
where $n_{e}$ and $n_{H}$ are the electron and hydrogen
number densities. Using the mass loss parameters 
for $\sigma$ Ori AB in Table 1 and assuming a solar-abundance
constant-velocity wind with v(r) = v$_{\infty}$ we have
EM$_{wind}$(r$>$R$_{o}$) = 5.5 $\times$ 10$^{55}$/R$_{o}$ cm$^{-3}$,
where R$_{o}$ is in units of R$_{*}$ = 9 R$_{\odot}$.
By comparison, the total X-ray emission
measure determined from the best-fit 2T $apec$ model
is EM$_{X}$ = 1.8 $\times$ 10$^{54}$ cm$^{-3}$.
To compute this value, we have converted the normalization
factors ($norm$) for both temperature components in Table 3 to 
equivalent emission measures and summed them, using
the conversion EM = 10$^{14}$4$\pi$d$_{cm}^2$$norm$,
where d$_{cm}$ is the stellar distance in cm. Adopting
R$_{o}$ = 1.5 as a representative example gives an inferred
ratio of X-ray EM to wind EM of $\approx$5\%.
Thus, the emission measure of the X-ray emitting plasma
relative to that of the wind is low.

The grating spectra provide clues  that a significant
fraction of the X-ray emission originates 
close to the star in the region where the wind is still 
accelerating. The relatively narrow lines and 
absence of any significant centroid
shifts are good evidence that the bulk of the X-ray
plasma is slowly moving and does not form at 
large radii where the wind
has reached terminal speed. Also, the very weak 
(formally undetected) forbidden lines in He-like triplets 
imply line formation close to the star.
As already noted, the most stringent constraint on
the location of the X-ray emitting plasma comes from
the Mg XI  triplet, for which we infer an upper
limit on the line formation radius r $<$ 1.8 R$_{*}$
(Fig. 22). Assuming a standard wind acceleration 
law v(r) = v$_{\infty}$[1 - (R$_{*}$/r)]$^{\beta}$ 
and v$_{\infty}$ = 1060 km s$^{-1}$ (Table 1),
the above radius is within the acceleration zone
for typical hot star wind values $\beta$ $\approx$ 
0.5 - 1.0. A recent analysis of He-like triplet
line ratios in O stars by Leutenegger et al. (2006)
concludes that minimum line formation radii are
typically 1.25 $<$ r/R$_{*}$ $<$ 1.67 and that
these small radii do not preclude the formation of
strong shocks.

A key question that remains to be answered is whether
numerical simulations of the line-driven instability mechanism
can reproduce the narrow line widths observed for
$\sigma$ Ori AB and some other late O and early B stars.
As already noted (Sec. 4.2) the line widths observed  for 
$\sigma$ Ori AB have HWHM values equivalent to about one-fourth
v$_{\infty}$. But larger widths would be in better accord
with numerical simulations, which tend to show shock onset
radii of about 1.5 R$_{*}$ and velocities of shock-heated material 
that are an appreciable fraction of v$_{\infty}$
(e.g. Feldmeier, Puls, \& Pauldrach 1997; Runacres \& Owocki 2002).
Even so, it should be kept in mind that predicted line properties 
from simulations are  sensitive to assumptions about wind properties
and the wind velocity profile close to the star, which are
uncertain for OB stars in general. A more detailed discussion of
the issues related to line widths and profiles in non-magnetic OB stars
in the framework of radiative wind shock models will be presented
in a future paper  (D. Cohen et al., in preparation).

\subsection{Magnetically-Confined Wind Shocks}

As already mentioned in the discussion of 
$\sigma$ Ori E (Sec. 3.7.2), a sufficiently strong surface
magnetic field can channel the wind of a hot star
into two streams, producing X-ray emission from
a  magnetically-confined
wind shock near the magnetic equator. To our knowledge,
there  have been no published reports of a magnetic field detection 
on $\sigma$ Ori AB. But,  an upper limit on its  line-strength 
weighted magnetic field  of
B$_{l}$ $\leq$ $-$140 ($\pm$260) G. was obtained 
in the longer of the two observations reported by
Landstreet (1982). Since the upper limit on
B$_{l}$ has a large uncertainty and this star is
young with a rather uncertain age 
 (Table 1 of Caballero 2007), it could still retain
a weak undetected fossil magnetic field. We thus briefly consider
the MCWS mechanism.

Using the relation for the wind confinement
parameter $\eta$  (Sec. 3.7.2)
and the adopted stellar parameters for
$\sigma$ Ori A (Table 1), the field strength
for critical confinement ($\eta$ = 1) is
B$_{eq}$ $\approx$ 70 G.  Thus, even
a relatively weak  field could perturb the
wind of $\sigma$ Ori A but a field of this strength
would be difficult to detect in a hot star.

Analytic expressions for the X-ray temperature
and luminosity of a magnetically-confined 
wind shock were derived by BM97a.
The predicted temperature at the shock front
is T$_{shock}$ = 1.13 $\times$ 10$^5$v$_{100}^2$ K,
where v$_{100}$ is the pre-shock wind speed
in units of 100 km s$^{-1}$. In order to
reconcile the MCWS picture with the relatively
cool emission at T $\sim$ 3.5 MK 
observed for $\sigma$ Ori AB, one would need to postulate
a pre-shock wind speed of v $\sim$ 550 km s$^{-1}$.
This is about half the terminal wind speed and
would require that the closed field lines exist
only in the wind acceleration zone near the star
(r $\ltsimeq$ 2 R$_{*}$ for values of
$\beta$ typical of hot star winds).
Intuitively, this is what one would expect in the
weak confinement regime.

The predicted X-ray luminosity from a 
magnetically-confined wind shock in the BM97a
model is L$_{x}$ $\approx$ 
2.6 $\times$ 10$^{30}$B$_{kG}^{0.4}$~$\xi$ ergs s$^{-1}$.
Here,  B$_{kG}$ is the field strength in kG and
$\xi$ is a free parameter of the theory that depends
on $\dot{M}$ and v$_{\infty}$. For critical 
confinement (B $\sim$ 70 G.), this relation
requires $\xi$ $\sim$ 35 to account for the
observed L$_{x}$ (Table 1). This value of
$\xi$  is within the (broad) range allowed by the 
theory and the adopted mass-loss parameters
of $\sigma$ Ori AB.

Based on the above comparisons, we conclude that the
MCWS model could plausibly account for the observed
X-ray properties of  $\sigma$ Ori AB if the magnetic field is
weak and only able to close the field lines near the star.
Detailed MCWS simulations would be needed to test this idea.
The justification for invoking the  MCWS model is not strong
in the absence of a magnetic field detection or any documented
H$\alpha$ or UV wind line modulation.

\subsection{Colliding Wind Shocks}

The presence of a B0.5V companion at a separation
0f 0.$''$25 from $\sigma$ Ori A raises the possibility
that X-rays could be produced in a colliding wind shock
region between the two stars. However, the predicted
maximum X-ray temperature for a colliding wind
shock in the case of solar abundances is 
kT$_{cw}$ $\approx$ 1.17 (v$_{\perp,1000}$)$^2$ keV,
where v$_{\perp,1000}$ is the velocity of the
wind component perpendicular to the shock 
interface in units of 1000 km s$^{-1}$
(Stevens, Blondin, \& Pollock 1992). Setting
v$_{\perp}$ equal to the terminal wind
speed of $\sigma$ Ori AB gives
kT$_{cw}$ $\approx$ 1.3 keV, much higher than observed.

Furthermore, wind momentum balance considerations
predict that the shock contact discontinuity
along the line-of-centers between the binary
components will lie at least
$\sim$1000 R$_{O-star}$ from $\sigma$ Ori A.
This equates to about half the distance between the
A and B components. To arrive at this estimate,
we have  assumed that the winds have reached
terminal speeds and have used eq. [1] of
Stevens, Blondin, \& Pollock (1992), along
with the mass-loss parameters in Table 1 and
the assumed B0.5V star wind parameters
log $\dot{M}_{B-star}$ $\sim$ $-$7.4 to $-$8.0
(M$_{\odot}$ yr$^{-1}$) and v$_{\infty,B-star}$
$\sim$ 2000 km s$^{-1}$ (Abbott 1982;
Cassinelli et al. 1994). X-ray formation at
such large distances is ruled out by the
weak or absent forbidden lines in He-like ions,
which for Mg XI demonstrate that the bulk of the X-rays 
are  formed within a few  radii of the O star.

Based on the above estimates, a colliding wind origin
for the X-rays seems very unlikely. But, before 
ruling out this mechanism it is important to note
that there have been several reports that $\sigma$ Ori A
is itself a double-lined spectroscopic binary
(e.g. Bolton 1974; Morrell \& Levato 1991), which if
true would make $\sigma$ Ori AB a triple system.
This  has also been 
mentioned  by Caballero (2008) as a possible means of
bringing the  derived distance of $\sigma$ Ori AB into better
agreement with that of the cluster.

If the A component
is indeed a spectroscopic binary then the wind of the more
massive primary could be shocking onto the wind of
the spectroscopic companion. From above, the wind speed needed to
produce the kT $\approx$ 0.3 keV plasma detected
by {\em Chandra} is v$_{wind}$ $\approx$
500 km s$^{-1}$. This would imply that the
winds are shocking at sub-terminal speeds.
For a $\beta$ = 1 velocity law, this speed
could be achieved at r $\approx$ 1.9 R$_{*}$
and for $\beta$ = 0.8 the computed radius
is r $\approx$ 1.6 R$_{*}$. Considering our
imprecise knowledge of the wind velocity 
law close to the star, these values are compatible 
with the upper limit r $\leq$ 1.8 R$_{*}$ obtained 
from Mg XI line flux ratios (Sec. 4.4).

A close colliding wind system consisting of 
$\sigma$ Ori A and a spectroscopic companion
might resemble close spectroscopic binaries
such as  HD 150136. This massive  O3 $+$ O6V
SB2 system has a short-period  2.66 day orbit
(Niemela \& Gamen 2005). A sensitive {\em Chandra}
HETG observation detected HD 150136 but 
the O3 $+$ O6V SB2 system could not be spatially
resolved (Skinner et al. 2005). The X-ray
spectrum of HD 150136 is soft with a 
characteristic temperature  kT $\approx$ 0.3 keV,
remarkably similar to that detected
here in $\sigma$ Ori AB.  Because of the close
separation of the two components in HD 150136,
the winds would be shocking at sub-terminal
speeds and radiative cooling in the shock
region is important (Skinner et al. 2005).
Thus, the predicted temperature for a colliding
wind shock is lower than for an equivalent 
adiabatic system at wide separation.
If the analogy between the 
O3 $+$ O6V spectroscopic binary  HD 150136 and 
$\sigma$ Ori A is indeed relevant, then an early B 
star may be playing the role of the spectroscopic companion 
to $\sigma$ Ori A (Caballero 2008).

If $\sigma$ Ori  A does have a spectroscopic companion,
then orbitally-modulated X-ray emission could be present
if the X-rays form in  a colliding wind shock. But,
the presence of orbital X-ray modulation depends critically
on the orbital eccentricity and inclination, for which no
information is currently available. We see no clear evidence for
X-ray modulation in the  $\sim$1 day {\em Chandra} light
curve (Fig. 4). More specific information on the properties of the
putative spectroscopic companion and its orbit would be 
needed to make further comparisons between colliding wind
shock theory and the X-ray data.

\subsection{Coronal Models and Late-Type Companions}

The existence of a thin X-ray emitting coronal zone 
at the base of the wind was  invoked by Cassinelli \& Olson (1979)
to explain strong O VI, N V, and C IV lines
in some OB supergiants and Of stars. Waldron \& Cassinelli (2001)
have also proposed a composite model involving both wind shocks
and confinement of dense plasma in magnetic loops close to the star
to  explain the broad  emission lines detected in the O9.7 Ib star 
$\zeta$ Ori, but Cohen et al. (2006) arrive at  a different 
interpretation that does not invoke magnetic confinement. 
Magnetic loop confinement close to the star strongly resembles 
coronal models for late-type stars.

But the
coronal model has not gained wide acceptance
due to the lack of a solid theoretical basis
for explaining how strong outer convection zones
would be produced in hot stars. Such convection
zones are thought to be essential for generating
self-sustained magnetic fields in the Sun and 
other late-type stars. Even so, the detection of
strong $\sim$kG magnetic fields in some very young O-type
stars such as  $\theta^1$ Ori C (Donati et al. 2002) 
does justify a brief
examination of the possibility of coronal emission,
either from $\sigma$ Ori AB or an unseen late-type
companion. 

Because of the young age of $\sigma$ Ori AB
($\sim$3 Myr), a coeval unseen  late-type companion
would likely be a T Tauri star and thus a potential
X-ray emitter. However, adaptive optics H-band images
with FWHM $\approx$ 0.$''$2 - 0.$''$4 (Caballero 2005) 
have only detected  IRS1, which at a separation of
$\approx$3$''$ is also revealed by {\em Chandra}. 
The best adaptive optics resolution of 
FWHM $\approx$ 0.$''$2 probes down to a linear
separation of $\sim$70 AU, so any as yet 
undetected T Tauri star would presumably  lie
very close to $\sigma$ Ori A. 

Most T Tauri stars detected in the deep {\em Chandra}
Orion COUP survey show a hot X-ray 
component at kT $>$ 1 keV (Preibisch et al. 2005).
We do not detect such a component in the X-ray
spectrum of  $\sigma$ Ori AB, so the observed
temperature structure provides no hint for
a T Tauri companion. 
Impulsive X-ray flaring or moderate amplitude
variability on short timescales is often (but not
always)   detected in TTS but no such variability
has so far been detected in $\sigma$ Ori AB. 
Finally, at the spectral resolution of {\em Chandra's}
HETG, emission lines of coronal late-type stars such
as Capella (Ness et al. 2003) and of TTS
such as SU Aur (Smith et al. 2005) and 
HDE 245059 (Baldovin Saavedra et al. 2007) show 
narrow lines with no broadening beyond
thermal widths, in contrast to the broadened
lines detected here in $\sigma$ Ori AB.
Thus, we find no strong
reason to invoke a coronal interpretation for 
the X-ray emission of $\sigma$ Ori AB.

\section{Conclusions}

The main conclusions of this study are the following:

\begin{enumerate}

\item {\em Chandra} detected 42 X-ray sources on the ACIS-S3
      CCD in the  central region of the $\sigma$ Ori cluster.
      Three are known high-mass OB stars and the remainder have
      X-ray and near-IR properties consistent with 
      magnetically-active K or M stars. Nineteen of the 
      X-ray sources have K or M spectral types determined
      from previous studies, and about half of these 
      show near-IR excesses based on {\em 2MASS} colors.

\item The X-ray emission associated with IRS1 is variable
      and shows a high-temperature (kT $\gtsimeq$ 1 keV)
      component. Its X-ray properties and  elongated
      mid-IR morphology  suggest that it is
      a low-mass magnetically-active T Tauri star whose
      disk/envelope is  being photo-evaporated by 
      by $\sigma$ Ori AB.

\item The X-ray emission of $\sigma$ Ori E is variable and
      a hard component (kT $\approx$ 2.6 keV) is detected, 
      as typical of magnetically-active stars. 
      Its ratio  L$_{x}$/L$_{*}$ is unusually high for a B2V 
      star assuming the nominal distance of 640 pc, suggesting
      that it could lie closer. Numerical magnetosphere models 
      based on the magnetically confined wind picture 
      predict hot plasma at 
      temperatures  similar to (but slightly higher than) observed
      and can also explain many aspects of the variability of 
      this star (Townsend et al. 2005, 2007). Future numerical
      models offer the potential to provide estimates of X-ray
      luminosity and X-ray absorption that
      can be directly compared with observations.

\item The X-ray emission of the OB binary system $\sigma$ Ori AB is
      exceptional, being  the most luminous X-ray source in 
      the field and  the softest.
      Its low X-ray temperature  (kT $\approx$ 0.3 keV)
      and moderately-broadened emission lines (HWHM $\approx$
      264 km s$^{-1}$)  clearly  argue against a coronal 
      interpretation and point   to a shock origin. Overall,
      its X-ray properties are quite similar to  some other 
      late O and early B stars for which X-ray grating spectra 
      show relatively cool 
      plasma and slightly broadened unshifted symmetric emission lines.

\item  Radiative wind shock predictions for an optically thin wind
       are consistent with the cool plasma and symmetric unshifted
       emission lines of  $\sigma$ Ori AB. However, the rather narrow
       line widths and low f/i ratios imply X-ray formation in slowly-moving
       plasma in the wind acceleration zone close to the star.
       It remains to be seen whether future  numerical simulations of the
       line-driven instability mechanism can reproduce the narrow lines.

\item  Other possible emission mechanisms for $\sigma$ Ori AB that are 
       not yet totally ruled out but remain speculative  are:
       (i) a magnetically confined wind shock in the weak-field
       limit, and (ii) a sub-terminal speed  colliding  wind system if 
       $\sigma$ Ori A does have a spectroscopic companion.  
       The MCWS scenario lacks solid observational support in the absence
       of any evidence for a magnetic field in  $\sigma$ Ori AB.
       More complete information about the properties of the putative
       spectroscopic companion and its orbit will be needed to make
       meaningful comparisons of colliding wind theory  with observations.

\end{enumerate}

\acknowledgments

We thank Ms. Victoria Swisher for assistance with data
analysis and Dr. Stephen Drake for providing information on
{\em VLA} observations of $\sigma$ Ori.
This research was supported by   grants 
GO5-6009X (SAO) and NNG05GE69G (NASA GSFC).

\clearpage

% Now comes the reference list.  In this document, we used \cite to call
% out citations, so we must use \bibitem in the reference list, which
% means we use the LaTeX thebibliography environment.  Please note that
% \begin{thebibliography} is followed by a null argument.  If you forget
% this, mayhem ensues, and LaTeX will say "Perhaps a missing item?" when
% you run it.  Do not call us, do not send mail when this happens.  Put
% the silly {} after the \begin{thebibliography}.
%
% Each reference has a \bibitem command to define the citation format
% to be placed in the text (in []) and the symbolic tag used for 
% cross referencing (in {}).
%
% See sample1.tex, or the AASTeX guide, for an alternative to the \cite-
% \bibitem command.

\clearpage

% TABLE1 : Sigma Ori Properties
\begin{deluxetable}{llll}
%%\rotate
\tabletypesize{\scriptsize}
\tablewidth{0pt}
\tablecaption{ $\sigma$ Ori AB Properties\tablenotemark{a}}
\tablehead{
\colhead{}      &
\colhead{}     &
\colhead{Reference}    & 
}
\startdata
Spectral Type &  O9.5V $+$ B0.5V &  (1)  \\
Distance (pc)\tablenotemark{b} &  352  $^{+166}_{-85}$    &  (2)   \\
V (mag)       &  3.80 - 3.81     &  (3),(4),(5)    \\
E(B$-$V) (mag)& 0.05 - 0.07\tablenotemark{c} &  (3),(4),(5)   \\
T$_{eff}$ (K)& 33,000            &  (6)  \\
R$_{*}$ (R$_{\odot}$) & 9        &  (6)  \\  
log $\dot{M}$ (M$_{\odot}$ yr$^{-1}$) & $-$7.1 & (6)   \\
v$_{\infty}$ (km s$^{-1}$)       &  1060      & (6)   \\
log [L$_{*}$/L$_{\odot}$]        & 4.9        & (6)  \\
log L$_{wind}$ (ergs s$^{-1}$)\tablenotemark{d} & 34.4   &  (6) \\
log L$_{x}$ (ergs s$^{-1}$) & 31.5   &  (7) \\
log [L$_{x}$/L$_{*}$]       & $-$7.0 &  (7)  \\

\enddata
\tablenotetext{a}{
Notes:~$\sigma$ Ori = HD 37468.  L$_{x}$ is the unabsorbed value
            in the 0.5 - 7 keV range. 
            Refs. (1) Edwards 1976; 
           (2) {\em Hipparcos} (Perryman et al. 1997); 
           (3) Lee 1968; 
           (4) Savage et al. 1985; 
           (5) Vogt 1976; 
           (6) Howarth \& Prinja 1989;
           (7) this work (0.5 - 7 keV).}
\tablenotetext{b}{We adopt d = 352 pc in this work. A 
 distance d = 334$^{+25}_{-22}$ pc was determined by
 Caballero (2008), who also noted that  d $\sim$ 385 pc is possible if
 the system is a triple.}
\tablenotetext{c}{Adopting A$_{\rm V}$ $\approx$ 3E(B$-$V) =
0.15 - 0.21 mag gives N$_{\rm H}$ =
(3.3 - 4.7) $\times$ 10$^{20}$ cm$^{-2}$ using the 
Gorenstein (1975) conversion.}
\tablenotetext{d}{L$_{wind}$ = $\dot{M}$v$_{\infty}^2$/2.}

\end{deluxetable}

\clearpage

% TABLE2 Chandra X-ray Sources (S3) 
\begin{deluxetable}{lllllll}
%%\rotate
\tabletypesize{\scriptsize}
\tablewidth{0pt}
\tablecaption{Chandra X-ray Sources in the Vicinity of $\sigma$ Ori AB\tablenotemark{a} }
\tablehead{
\colhead{CXO}      &
\colhead{R.A.}     &
\colhead{Decl.}    & 
\colhead{Net Counts}   &
\colhead{$<$E$>$}   &
\colhead{K$_{s}$}  &
\colhead{Identification(offset)} \\
\colhead{No.   }    &
\colhead{(J2000)     }    &
\colhead{(J2000)     }    &
\colhead{(cts)  }   &
\colhead{(keV)       }   &
\colhead{(mag)}    &
\colhead{(arcsec)       }
}
\startdata
1  & 05 38 29.12   & -02 36 03.0   &   69$\pm$8(v.99)  & 1.58 & 11.69 &  2M053829.12-023602.7(0.30)\\
2  & 05 38 31.43   & -02 36 33.8   &    7$\pm$3        & 1.29 & 10.99 &  2M053831.41-023633.8(0.24)\\
3  & 05 38 31.57   & -02 35 14.7   &   32$\pm$6        & 2.07 & 10.35 &  2M053831.58-023514.9(0.24)\\
4  & 05 38 32.83   & -02 35 39.4   &  147$\pm$12       & 1.35 & 10.73 &  2M053832.84-023539.2(0.30)\\
5  & 05 38 33.34   & -02 36 18.3   &   21$\pm$5        & 1.41 & 11.11 &  2M053833.36-023617.6(0.72)\\
6  & 05 38 34.04   & -02 36 37.2   &    8$\pm$3        & 1.10 & 11.08 &  2M053834.06-023637.5(0.42)\\
7  & 05 38 34.31   & -02 35 00.2   &   50$\pm$7        & 1.29 & 10.35 &  2M053834.31-023500.1(0.12)\\
8  & 05 38 35.21   & -02 34 38.1   &  369$\pm$19(v.95) & 3.01 & ...   &  M07:053835.22-023438.0(0.18)  \\
9  & 05 38 38.22   & -02 36 38.6   &  390$\pm$20(v.99) & 1.79 & 10.31 &  2M053838.22-023638.4(0.18)\\
10 & 05 38 38.48   & -02 34 55.2   &  697$\pm$26(v.97) & 1.60 &  9.12 &  2M053838.49-023455.0(0.24)\\
11 & 05 38 39.72   & -02 40 19.2   &   12$\pm$4(v.96)  & 3.09 & 12.88 &  2M053839.73-024019.7(0.54) \\
12 & 05 38 41.21   & -02 37 37.4   &   29$\pm$5        & 3.40 & 14.04 &  2M053841.23-023737.7(0.48)\\
13 & 05 38 41.28   & -02 37 22.7   &  323$\pm$18       & 1.64 & 10.59 &  2M053841.29-023722.6(0.24)\\
14 & 05 38 41.35   & -02 36 44.6   &   14$\pm$4        & 1.13 & 12.08 &  2M053841.35-023644.5(0.12)\\
15 & 05 38 42.27   & -02 37 14.5   &    9$\pm$3        & 2.56 & 10.77 &  2M053842.28-023714.7(0.30)\\
16 & 05 38 43.01   & -02 36 14.7   &    6$\pm$3        & 1.93 & 10.63 &  2M053843.02-023614.6(0.18)\\
17 & 05 38 43.85   & -02 37 07.3   &   16$\pm$4        & 1.38 & 11.77 &  2M053843.87-023706.9(0.54)\\
18 & 05 38 44.22   & -02 40 20.0   &  158$\pm$13       & 1.52 & 10.44 &  2M053844.23-024019.7(0.30)\\
19 & 05 38 44.76   & -02 36 00.3   & 2514$\pm$50       & 0.86 &  4.49 &  2M053844.76-023600.2(0.12); $\sigma$ Ori AB\\
20 & 05 38 44.83   & -02 35 57.4   &   43$\pm$7(v.95)  & 1.37 & ...   &  $\sigma$ Ori IRS1 \\
21 & 05 38 45.35   & -02 41 59.6   &    50$\pm$5       & 1.69 & 11.04 &  2M053845.37-024159.4(0.42)\\
22\tablenotemark{c} & 05 38 46.82   & -02 36 43.5      &    6$\pm$2     & 0.95 & 12.35 &  2M053846.85-023643.5(0.48)\\
23 & 05 38 47.19   & -02 35 40.5   &  454$\pm$21(v.98) & 1.56 &  6.95 &  2M053847.20-023540.5(0.18); $\sigma$ Ori E\\
24 & 05 38 47.46   & -02 35 25.4   &   87$\pm$9        & 1.64 & 10.72 &  2M053847.46-023525.3(0.18)\\
25 & 05 38 47.89   & -02 37 19.6   &  188$\pm$14(v.99) & 2.20 & 10.78 &  2M053847.92-023719.2(0.54)\\
26 & 05 38 48.27   & -02 36 41.2   &   42$\pm$7        & 1.35 & 11.14 &  2M053848.29-023641.0(0.30)\\
27 & 05 38 48.68   & -02 36 16.4   &  106$\pm$10(v.95) & 1.35 & 11.17 &  2M053848.68-023616.2(0.18)\\
28 & 05 38 49.12   & -02 41 23.9   &    22$\pm$6       & 1.62 & ...  &  S04:053849.14-024124.8 (0.96)\\
29 & 05 38 49.17   & -02 38 22.5   &  112$\pm$11       & 1.32 & 10.51 &  2M053849.17-023822.2(0.24)\\
30 & 05 38 49.69   & -02 34 52.6   &    8$\pm$3        & 1.22 & 12.09 &  2M053849.70-023452.6(0.24)\\
31 & 05 38 50.02   & -02 37 35.5   &   15$\pm$4        & 1.64 & 12.10 &  2M053850.03-023735.5(0.18)\\
32\tablenotemark{b,c} & 05 38 50.16   & -02 36 54.6    &    5$\pm$2 &  4.06  & ...    & ... \\
33 & 05 38 51.43   & -02 36 20.7   &   53$\pm$7        & 1.86 & 11.55 &  2M053851.45-023620.6(0.36)\\
34 & 05 38 51.75   & -02 36 03.3   &    6$\pm$3        & 1.56 & 12.03 &  2M053851.74-023603.3(0.18)\\
35 & 05 38 51.85   & -02 33 33.0   &   17$\pm$4        & 2.94 & ...   & ... \\
36 & 05 38 53.03   & -02 38 53.7   &   12$\pm$4        & 1.50 & 10.83 &  2M053853.07-023853.6(0.61) \\
37 & 05 38 53.37   & -02 33 23.2   &  863$\pm$29(v.99) & 1.67 &  9.73 &  2M053853.37-023323.0(0.24)\\
38 & 05 38 58.24   & -02 38 51.5   &   25$\pm$5        & 2.67 & ...    & ... \\
39 & 05 38 59.03   & -02 34 13.2   &   13$\pm$4        & 2.73 & ...    & ... \\
40 & 05 38 59.51   & -02 35 28.6   &   12$\pm$4        & 3.06   & ...    & ... \\
41 & 05 39 00.53   & -02 39 38.7   &  94$\pm$10        &   1.66   & 11.11  &  2M053900.53-023939.0(0.30)\\
42 & 05 39 01.47   & -02 38 56.5   &  342$\pm$19       &  1.42   &  8.09  &  2M053901.50-023856.4(0.36); HD 37525\tablenotemark{d}\\

\enddata
\tablenotetext{a}{
Notes:~X-ray data are from CCD7 (ACIS chip S3) using events in the 0.5 - 7 keV range.
Tabulated quantities are: running source number, X-ray position (R.A., decl.), net counts and 
net counts error from {\em wavdetect} (accumulated in a 90,967 s exposure, rounded to the nearest integer,
background subtracted and PSF-corrected), mean photon energy $<$E$>$,
K$_{s}$ magnitude of near-IR 2MASS counterpart, and 2MASS (2M) or optical (S04 = Sherry, Walter, 
\& Wolk 2004; M07 = Mayne et al. 2007) candidate counterpart identification
within a 2$''$ search radius. The offset (arc seconds) between the X-ray and counterpart position is given 
in parentheses. A (v) following Net Counts error indicates that the source is likely variable as indicated
by a variability probability P$_{var}$ $\geq$ 0.95 determined from the Kolmogorov-Smirnov (KS) statisitic.
The number following v is the KS variability probability, i.e. v.99 indicates a variability 
probability P$_{var}$ = 0.99.
All sources were confirmed to be present in the archived  
98 ksec {\em Chandra}  HRC-I image (Obsid 2560).}

\tablenotetext{b}{
 Probable {\em XMM-Newton} counterpart
 is source NX 99 in Table B.1 of Franciosini et al. (2006). High value of $<$E$>$ suggests
 possible extragalactic background source. }

\tablenotetext{c}{ 
  Low-significance {\em wavdetect} detection (2. $<$ significance $<$ 3.).}

\tablenotetext{d}{
Double star with a B5V primary and a companion at separation 
0.$''$45 (Caballero 2005).}

\end{deluxetable}

%%TABLE3 - CCD Spectral fit results: Sig Ori

\clearpage
\begin{deluxetable}{lllllllc}
%%\rotate
\tabletypesize{\tiny}
\tablewidth{0pt}
\tablecaption{Spectral Fits of Zero-Order ACIS-S3 CCD Spectra\tablenotemark{a} }
\tablehead{
\colhead{CXO}      &
\colhead{N$_{\rm H}$}     &
\colhead{kT$_{1}$}    & 
\colhead{kT$_{2}$}   &
\colhead{norm$_{1}$}   &
\colhead{norm$_{2}$}  &
\colhead{$\chi^2$/dof ($\chi^2_{red}$)} & 
\colhead{Flux}    \\
\colhead{No.}    &
\colhead{(10$^{20}$ cm$^{-2}$)}  & 
\colhead{(keV)}  & 
\colhead{(keV)}  &
\colhead{ (10$^{-5}$)}  & 
\colhead{(10$^{-5}$)}  & 
\colhead{   }    &
\colhead{(10$^{-13}$ ergs cm$^{-2}$ s$^{-1}$)} 
}
\startdata
4 & \{3.0\}           & 0.37 [0.23-0.51] & 1.4 [1.2-1.7] & 1.71 [0.67-2.27] & 3.32 [2.38-3.99] & 5.6/5 (1.12)      & 0.70 (0.76) \\
8 &  5.0 [3.8-8.8]    & 0.69 [0.47-2.16] & 19.8 [8.1-...]& 3.49 [0.10-13.6] & 14.1 [12.4-17.4] & 32.3/31 (1.04)\tablenotemark{b}      & 1.97 (3.07) \\
9 &  1.5 [0.0-4.5]    & 0.61 [0.35-0.78] & 2.5 [2.1-2.9] & 2.14 [1.44-2.79] & 10.4 [9.00-12.0] & 31.2/31 (1.01)    & 1.87 (1.93) \\
10 & 2.5 [1.2-5.0]    & 0.70 [0.61-0.76] & 2.4 [2.1-2.8] & 5.81 [4.83-7.10] & 14.4 [12.0-16.3] & 72.8/52 (1.40)    & 3.19 (3.37) \\
13 &  \{3.0\}         & 0.74 [0.63-0.84] & 3.1 [2.3-4.8] & 2.58 [1.52-3.37] & 7.19 [5.53-8.52] & 11.4/11 (1.04)    & 1.55 (1.65) \\
18 &  \{3.0\}         & 0.37 [0.26-0.59] & 2.0 [1.6-2.5] & 1.89 [0.54-2.56] & 3.73 [2.96-4.78] & 13.1/14 (0.94)    & 0.78 (0.84) \\
19\tablenotemark{c} &  \{3.0\}  & 0.30 [0.29-0.31] & ...           & 111. [105.-116.] & ...              & 76.5/49 (1.56)   & 15.4 (17.8) \\
19\tablenotemark{c} &  \{3.0\}  & 0.21 [0.18-0.24] & 0.47 [0.45-0.55] & 79.9 [69.0-90.9] & 41.4 [25.8-50.3] & 45.6/47 (0.97)& 16.6 (19.5) \\
20 &  \{3.0\}         & 0.61 [0.27-0.83] & 2.2 [1.2-...] & 0.59 [0.27-0.89] & 0.61 [0.10-1.11] & 1.2/6 (0.20)     & 0.21 (0.23) \\
23\tablenotemark{d} & \{3.0\}   & 0.72 [0.61-0.80] & 2.4 [2.0-3.0] & 4.43 [3.46-5.24] & 8.46 [6.91-10.4] & 28.6/24 (1.19)   & 2.09 (2.23) \\
25 &  \{3.0\}         & 3.20 [2.59-4.09] & ...           & 6.96 [6.07-7.86] &      ...         & 18.1/17 (1.06)   & 0.96 (1.00) \\
27 &  \{3.0\}         & 0.66 [0.49-0.82] & 2.2 [1.5-3.4] & 1.24 [0.81-1.62] & 1.79 [0.96-2.55] & 6.4/7 (0.91)     & 0.50 (0.54) \\
29 &  \{3.0\}         & 0.67 [0.47-0.82] & 1.6 [1.1-3.2] & 1.29 [0.67-1.89] & 1.71 [0.71-2.52] & 7.0/8 (0.88)     & 0.52 (0.56) \\
37 & 2.1 [0.9-4.1]    & 0.75 [0.65-0.83] & 2.3 [2.1-2.6] & 6.02 [4.42-7.06] & 24.9 [22.0-27.8] & 26.2/32 (0.82)   & 4.59 (4.80) \\
42 &  \{3.0\}         & 0.76 [0.68-0.83] & 2.6 [2.0-3.5] & 4.69 [3.65-5.49] & 5.56 [4.15-7.37] & 34.4/29 (1.19)   & 1.76 (1.89) \\
\enddata
\tablenotetext{a}{
Notes:~The CXO no. in column (1) refers to Table 2. 
Fits are based on one-temperature (1T) or two-temperature (2T) optically thin plasma
{\em apec} solar-abundance models of rebinned spectra in XSPEC (v. 12.3).  
Solar abundances are referenced to Anders \& Grevesse (1989).
The notation N$_{\rm H}$ = \{3.0\} means the absorption column density was held fixed at that
value during fitting.  Brackets enclose 90\% confidence intervals and an 
ellipsis means the algorithm used to compute the 90\% confidence limit did not converge.
The {\em  norm} is related to the volume emission measure (EM) by
EM $=$ 10$^{14}$4$\pi$d$_{cm}^2${\em norm}, where d$_{cm}$ is the distance (cm) to the source.
The total X-ray fluxes are the absorbed values in the 0.5-7 keV
range, followed  in parentheses by unabsorbed values. }
\tablenotetext{b}{An absorbed  power-law model with a photon power-law index $\alpha$ = $+$1.6 [1.4 - 1.8]
                  and N$_{\rm H}$ = 2.5e20 cm$^{-2}$ gives nearly identical fit statistics 
                  with $\chi^2$/dof = 34.7/33 and an unabsorbed flux 2.47e-13 ergs cm$^{-2}$ s$^{-1}$.}
%%\tablenotetext{c}{A variable abundance 1T {\em apec} fit converges to a global
%%                  abundance Z = 0.20 [0.14-0.28] Z$_{\odot}$ with kT$_{1}$ = 0.30 keV
%%                  ($\chi^2$/dof = 71.1/52). An analogous 2T {\em apec} fit converges to
%%                  Z = 0.25 [0.15-0.38] Z$_{\odot}$ with kT$_{1}$ = 0.30 keV and
%%                  kT$_{2}$ = 1.7 keV ($\chi^2$/dof = 59.2/50).}
\tablenotetext{c}{Estimated zero-order photon pileup is 5\% for $\sigma$ Ori AB. Events with energies
                  below 0.5 keV were excluded from the fit to avoid any possible contamination by soft UV/optical leaks.}
\tablenotetext{d}{A 3T $apec$ model gives an improved fit with N$_{\rm H}$ = 2.6 [1.1-10.4] $\times$ 10$^{20}$ cm$^{-2}$,
                  kT$_{1}$ = 0.24 [0.09-0.39] keV, kT$_{2}$ = 0.81 [0.75-1.05] keV, kT$_{3}$ = 2.6 [2.1-3.3] keV, 
                  norm$_{1}$ = 2.91 [1.00-12.6]e-5, norm$_{2}$ = 3.77 [2.05-5.42]e-5, norm$_{3}$ = 8.11 [6.73-9.76]e-5, 
                  $\chi^2$/dof = 18.6/21 ($\chi^2_{red}$ = 0.89), and Flux = 2.24 (2.41) $\times$ 10$^{-13}$
                  ergs cm$^{-2}$ s$^{-1}$.}
\end{deluxetable}

\clearpage

% TABLE4.TEX 
\begin{deluxetable}{llllllc}
%%\rotate
\tabletypesize{\footnotesize}
\tablewidth{0pt}
\tablecaption{$\sigma$ Ori AB Emission Line Properties\tablenotemark{a} }
\tablehead{
\colhead{Ion}      &
\colhead{$\lambda_{lab}$}     &
\colhead{$\lambda_{obs}$}    & 
\colhead{$\Delta\lambda$}   &
\colhead{Line Flux}   &
\colhead{HWHM}  &
\colhead{log T$_{max}$}  \\
\colhead{   }    &
\colhead{(\AA)     }    &
\colhead{(\AA)     }    &
\colhead{(m\AA)  }   &
\colhead{(10$^{-5}$ ph cm$^{-2}$ s$^{-1}$)       }   &
\colhead{(km s$^{-1}$)} &
\colhead{(K)}    
}
\startdata
Si XIII(r) & 6.6479  &  ...                     & ...   & 0.12 (0.01 - 0.22)   & \{300\} & 7.0       \\
Mg XI(r)   & 9.1687  &  9.165 (9.162 - 9.169)   & $-$4. & 0.42 (0.30 - 0.55)   & 250 (8 - 374) & 6.8 \\
Mg XI(i)   & 9.2312  &  ...                     & ...   & 0.41 (0.27 - 0.52)   & 250\tablenotemark{c} & 6.8        \\
Mg XI(f)   & 9.3143  &  ...                     & ...   & ...\tablenotemark{d} & ... & 6.8  \\ 
Ne X       &10.239   &  ...                     & ...   & 0.25 (0.09 - 0.42)   & \{300\} & 6.8   \\
Ne IX      &11.544   &  ...                     & ...   & 0.48 (0.23 - 0.74)   & \{300\} & 6.6       \\ 
Ne X\tablenotemark{b} &12.132   & 12.127 (12.124 - 12.130) & $-$5. & 1.64 (1.28 - 1.98)   & 390 (248 - 542) & 6.8 \\
Ne IX(r)   &13.447   & 13.446 (13.441 - 13.449) & $-$1. & 3.36 (2.80 - 4.03)   & 314 (216 - 381) & 6.6 \\
Ne IX(i)   &13.553   & 13.546 (13.543 - 13.548) & $-$7. & 3.53 (2.87 - 4.09)   & 314\tablenotemark{c} & 6.6        \\
Ne IX(f)   &13.699   & ...                      & ...   & ...\tablenotemark{d} & ... & 6.6  \\
Fe XVII    &15.014   & 15.013 (15.010 - 15.016) & $-$1. & 7.94 (6.69 - 9.14)   & 279 (195 - 355) & 6.7 \\
O VIII     &15.176   & ...                      & ...   & 1.11 (0.53 - 1.98)   & \{300\}  & 6.5      \\
Fe XIX     &15.198   & ...                      & ...   & 1.01 (0.55 - 2.24)   & \{300\}  & 6.9       \\
Fe XVII    &15.261   & 15.258 (15.251 - 15.266) & $-$3. & 3.09 (2.36 - 3.83)   & \{300\}  & 6.7 \\
O VIII     &16.006   & 16.009 (16.002 - 16.015) & $+$3. & 2.63 (1.57 - 3.60)   & 195 (0 - 405) & 6.5  \\
Fe XVII    &16.780   & 16.778 (16.772 - 16.785) & $-$2. & 4.98 (3.66 - 6.27)   & 155 (21 - 278) & 6.7 \\
Fe XVII    &17.051   & 17.049 (17.043 - 17.055) & $-$2. & 5.32 (4.15 - 6.63)   & 163 (92 - 240) & 6.7    \\
Fe XVII    &17.096   & 17.091 (17.086 - 17.096) & $-$5. & 6.91 (5.58 - 8.29)   & \{163\} & 6.7     \\
O VII      &18.627   & ...                      & ...   & 1.86 (0.25 - 3.46)   & \{300\} & 6.3       \\
O VIII\tablenotemark{b} &18.967   & 18.968 (18.964 - 18.972) & $+$1. & 22.0 (19.6 - 27.8)   & 330 (273 - 435) & 6.5 \\
O VII(r)   &21.602   & 21.608 (21.598 - 21.617) & $+$6. & 15.5 (10.7 - 21.8)   & 298 (97 - 507) & 6.3 \\
O VII(i)   &21.804   & 21.803 (21.796 - 21.810) & $-$1. & 22.4 (15.2 - 30.1)   & 298\tablenotemark{c} & 6.3     \\
O VII(f)   &22.098   & ...                      & ...   & ...\tablenotemark{d} & ... & 6.3   \\
\enddata
\tablenotetext{a}{
Notes:~X-ray data are from the background-subtracted HETG/MEG1 spectrum 
($+$1 and $-1$ orders combined), regrouped to a minimum
of 10 counts per bin. Parentheses enclose 1$\sigma$ intervals.
Tabulated quantities are: ion name (Ion) where r,i,f denote resonance, intercombination, and
forbidden lines,  laboratory wavelength of transition ($\lambda_{lab}$),
measured wavelength  ($\lambda_{obs}$), $\lambda_{obs}$ $-$ $\lambda_{lab}$ ($\Delta\lambda$),
observed (absorbed) continuum-subtracted line flux (Line Flux), line half-width at half-maximum
(HWHM), and maximum line power temperature (T$_{max}$).
An ellipsis means that $\lambda_{obs}$ was not measured due to faint or closely-spaced lines.
Curly braces enclose quantities that were held fixed during fitting.
The value of $\lambda_{obs}$ is subject to the $\pm$11 m\AA~ absolute wavelength 
calibration accuracy of HETG/MEG.}
\tablenotetext{b}{Blend.}
\tablenotetext{c}{The line width was allowed to vary during fitting but was kept equal to the
width of the resonance (r) line.}
\tablenotetext{d}{The forbidden line was undetected.}
\end{deluxetable}

\clearpage

% TABLE5.TEX
\begin{deluxetable}{lll}
\tablewidth{0pc}
\tabletypesize{\scriptsize}
\tablecaption{$\sigma$ Ori AB He-like Ion  Line Flux Ratios\tablenotemark{a} }
\tablehead{
\colhead{ } &
\colhead{Mg XI } &
\colhead{Ne IX }  \\
}
\startdata
$\lambda_{lab,r}$ (\AA~)  & 9.1687                  & 13.447\tablenotemark{b}    \nl
log T$_{max}$ (K)         & 6.8                     & 6.6                        \nl
$G$\tablenotemark{c}        & 1.21 [0.76 - 1.92]    & 0.99 [0.73 - 1.32]         \nl
$R$\tablenotemark{c}        & $\leq$0.17            & $\leq$0.15                 \nl 
$R_{0}$                   & 2.7                     & 3.6                        \nl
$\phi_{c}$ (s$^{-1}$)     & 4.86 $\times$ 10$^{4}$  & 7.73 $\times$ 10$^{3}$     \nl
$n_{c}$ (cm$^{-3}$)       & 6.2 $\times$ 10$^{12}$  & 6.4 $\times$ 10$^{11}$     \nl
r$_{fi}$/R$_{*}$          & $\leq$1.8            & $\leq$4.4                     \nl
\tablenotetext{a}{
The laboratory rest wavelength of the resonance line
$\lambda_{lab,r}$ and the maximum line power temperature
T$_{max}$ are from APED. 
The line flux ratios $G$ = (f $+$ i)/r and $R$ = f/i are from ISIS
spectral fits (see below).  
$R_{0}$ is the theoretical  limit of $R$  at T$_{max}$
for low photoexcitation and low density
(Ne IX from Pradhan 1982; Mg XI from Porquet et al. 2001).
The critical photoexcitation rate $\phi_{c}$ and critical
density $n_{c}$ are from BDT72.
The 90\% confidence upper limit on the line formation radius
is r$_{fi}$/R$_{*}$.
}
%% O VII from Porquet 2005).} 
\tablenotetext{b}{The Ne IX resonance line may be weakly contaminated by  
emission from  Fe XIX (13.462 \AA~).}
\tablenotetext{c}{Determined from simultaneous fits of MEG1$+$HEG1 spectra
with ISIS. The He-like complexes were fitted allowing $R$, $G$, and the metal
abundances to vary. ISIS applies the appropriate instrumental and
thermal broadening to each line. The best-fit abundances 
of Mg and Ne are consistent
with solar at the 90\% confidence level. The continuum level was 
determined by a global 2T $vapec$ fit of the zero-order spectrum. 
Square brackets enclose 90\% confidence intervals on $G$ and upper
limits on $R$  are 90\% confidence.}
\enddata
\end{deluxetable}

% And finally, we must deal with the figures.  

%% Uncomment the following line to skip the figures
%%\end{document}

\begin{figure}
\figurenum{1}
\epsscale{1.0}
\includegraphics*[width=8.5cm,angle=-90]{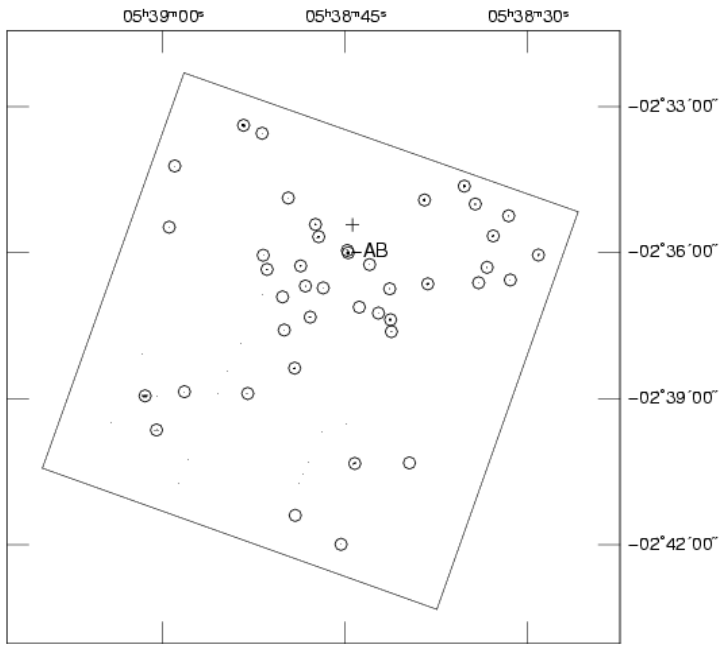}
\caption{Circles mark the positions of the 42 X-ray 
sources detected on the ACIS-S3 CCD as listed in Table 2.
A cross marks the {\em Chandra} aimpoint and the primary
target $\sigma$ Ori AB is identified.}
\end{figure}

\clearpage
\begin{figure}
\figurenum{2}
\epsscale{1.0}
\includegraphics*[width=8.5cm,angle=-90]{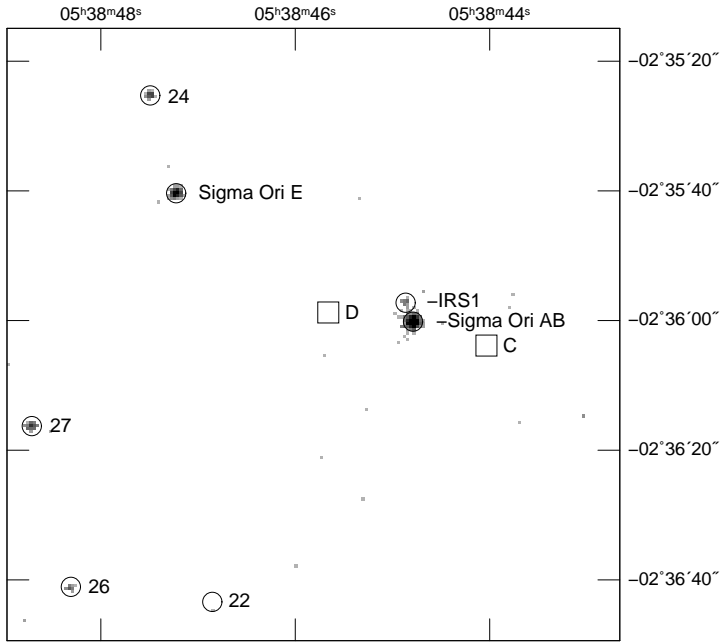}
\caption{Unsmoothed ACIS-S3 image (0.5 - 7 keV) of
 the region near $\sigma$ Ori AB.
 Circles mark ACIS-S3   detections and squares mark
 the optical positions of undetected stars $\sigma$ Ori C and 
 D. Numbered sources correspond to Table 2.}
\end{figure}

\clearpage
\begin{figure}
\figurenum{3}
\epsscale{1.0}
\includegraphics*[width=8.5cm,angle=-90]{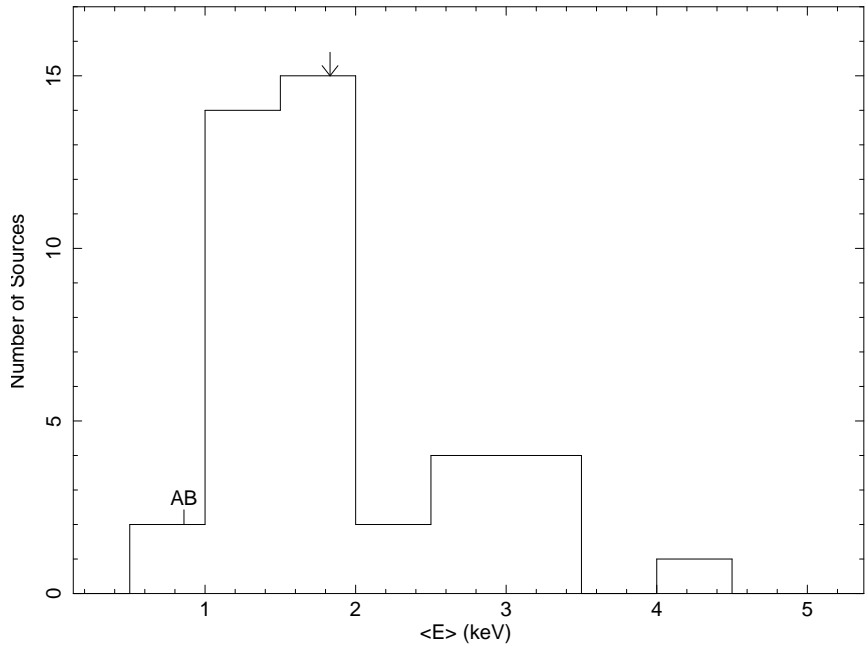}
\caption{Histogram of mean photon energies for X-ray sources
         on ACIS-S3. The value of $\sigma$ Ori AB is shown and
         the downward arrow marks the mean value.}
\end{figure}

\clearpage
\begin{figure}
\figurenum{4}
\epsscale{1.0}
\includegraphics*[width=8.5cm,angle=-90]{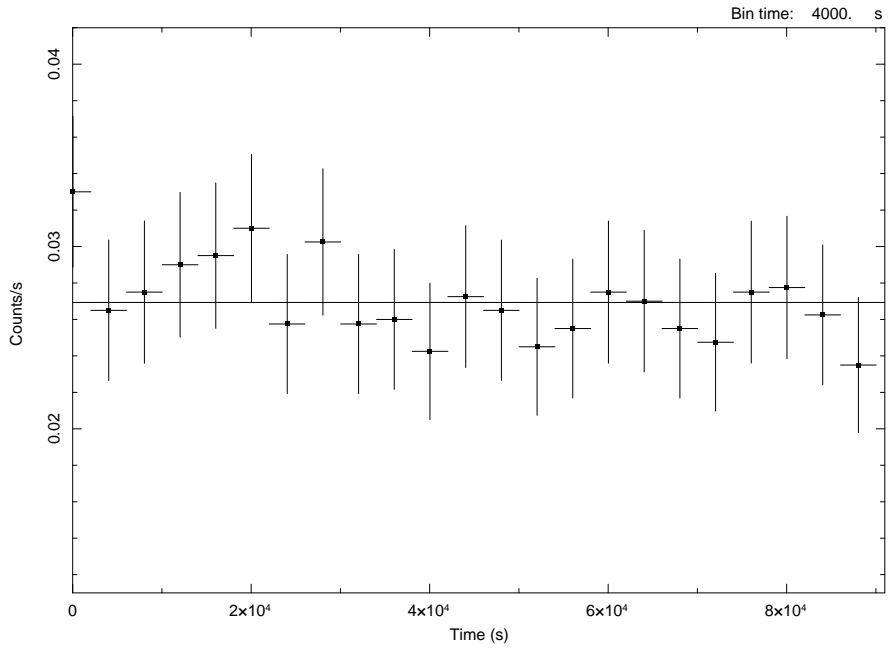}
\caption{ACIS-S3 light curve of $\sigma$ Ori AB (CXO  19) binned
 at 4000 s intervals using events in the  0.5 - 7 keV
 range and extraction radius R$_{e}$ = 1.$''$97.
The solid line is a best-fit constant count rate model.
Error bars are 1$\sigma$. The KS variability probability 
is P$_{var}$ = 0.935.
}
\end{figure}

\clearpage
\begin{figure}
\figurenum{5}
\epsscale{1.0}
\includegraphics*[width=8.5cm,angle=-90]{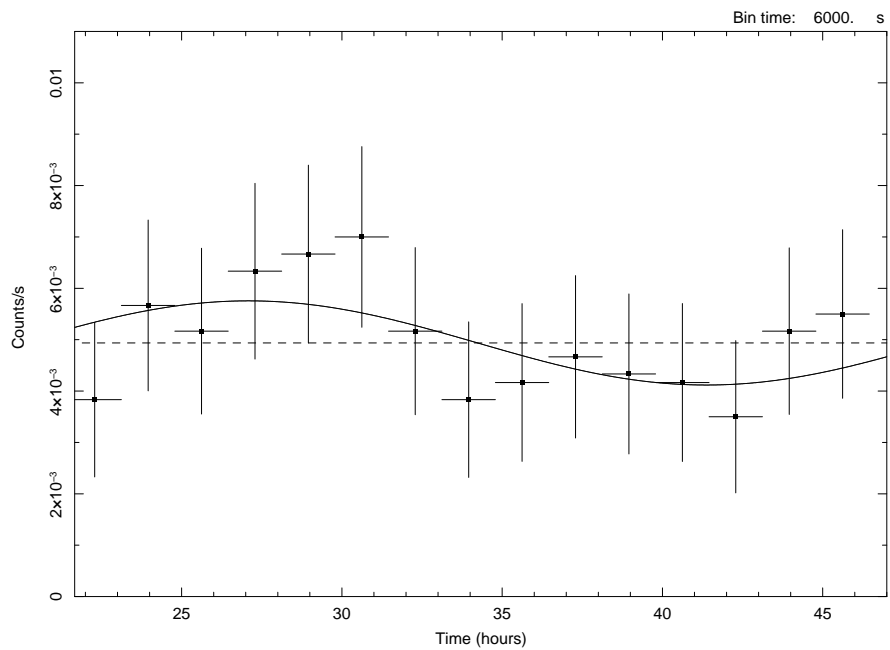}
\caption{ACIS-S3 light curve of $\sigma$ Ori E (CXO  23) binned
 at 6000 s intervals using events in the  0.5 - 7 keV
 range and extraction radius R$_{e}$ = 1.$''$97.
 Error bars are 1$\sigma$. A KS test gives a high  probability
 of variability P$_{var}$ = 0.986. The solid curve is a best fit
sinusoid model with a constant offset of 0.004937 counts/s
(dashed line) and the period fixed at the stellar rotation period 
P$_{rot}$ = 1.19084 d = 28.58 hours. The time axis 
shows the elapsed time in hours from start of the day.
The {\em Chandra} observation began at 20:59:02 UT on
2003 August 12 and spanned 1.053 d. }
\end{figure}

\clearpage
\begin{figure}
\figurenum{6}
\epsscale{1.0}
\includegraphics*[width=8.5cm,angle=-90]{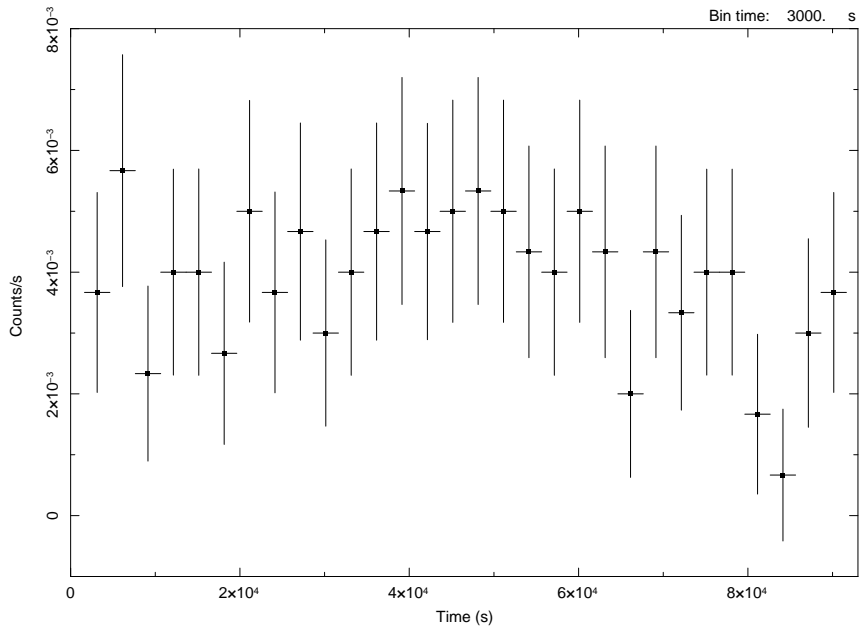}
\caption{ACIS-S3 light curve of the variable source
 CXO J053835.21$-$023438.1 (CXO  8)
 binned  at 3000  s intervals using events in the  0.5 - 7 keV
 range and extraction radius R$_{e}$  = 1.$''$97.
 Error bars are 1$\sigma$.
 No {\em 2MASS} counterpart was found but a faint V = 20.88 mag
 optical source lies at an offset of 0.$''$18 from the X-ray
 position (Mayne et al. 2007).  
}
\end{figure}

\clearpage
\begin{figure}
\figurenum{7}
\epsscale{1.0}
\includegraphics*[width=8.5cm,angle=-90]{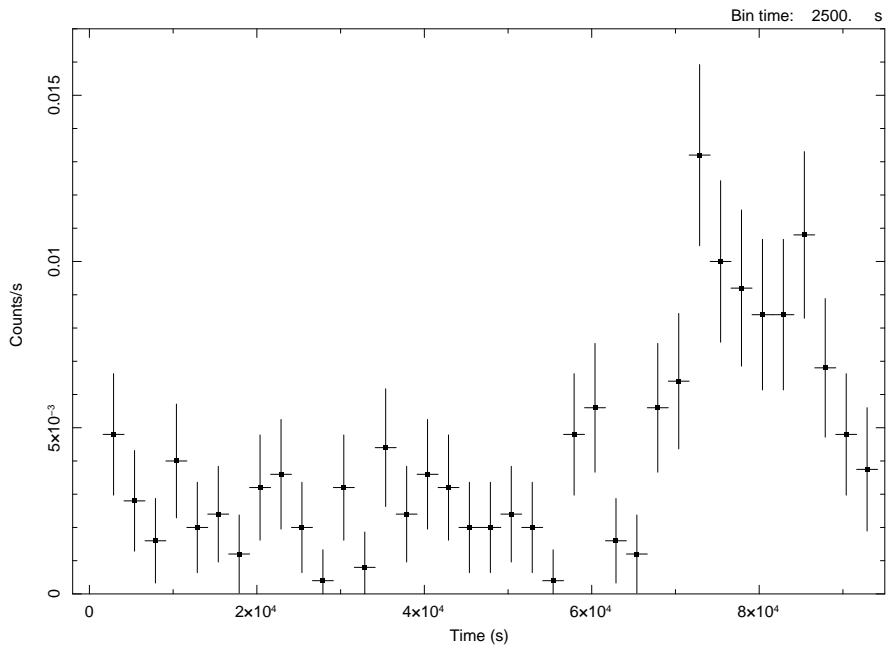}
\caption{ACIS-S3 light curve of CXO J053838.22$-$023638.6 (CXO  9)
 binned  at 2500  s intervals using events in the  0.5 - 7 keV
 range and extraction radius R$_{e}$ =  1.$''$97.
 Error bars are 1$\sigma$.
 This T Tauri star shows weak H$\alpha$ emission and was classified as 
 K8 by Zapatero Osorio (2002).}
\end{figure}

\clearpage
\begin{figure}
\figurenum{8}
\epsscale{1.0}
\includegraphics*[width=8.5cm,angle=-90]{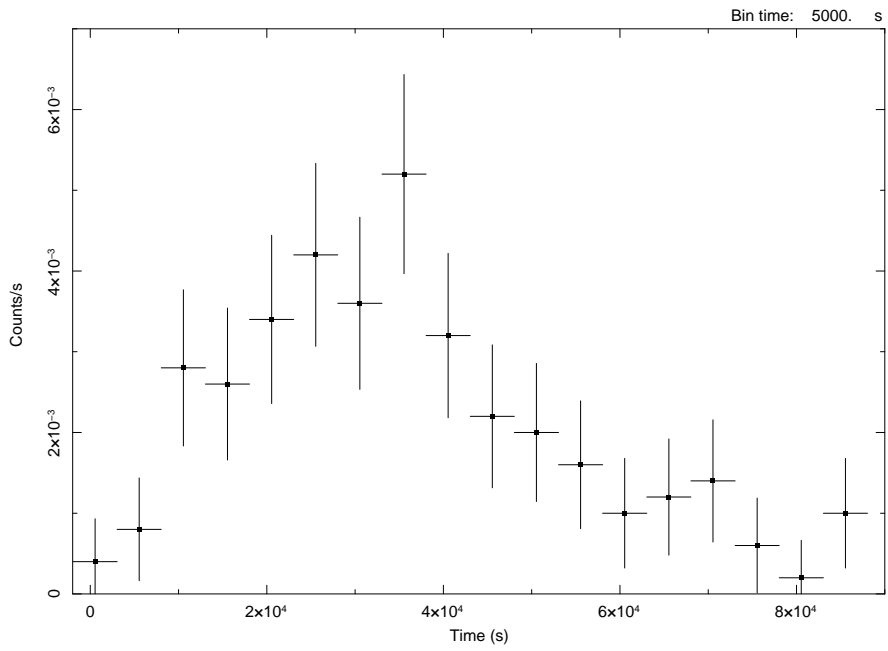}
\caption{ACIS-S3 light curve of CXO J053847.89$-$023719.6 (CXO 25)
 binned  at 5000  s intervals using events in the  0.5 - 7 keV
 range and extraction radius R$_{e}$  = 1.$''$97.
 Error bars are 1$\sigma$. 
 This star has been classified as M4 and shows a near-IR
 excess (Oliveira et al. 2006).}
\end{figure}

\clearpage
\begin{figure}
\figurenum{9}
\epsscale{1.0}
\includegraphics*[width=8.5cm,angle=-90]{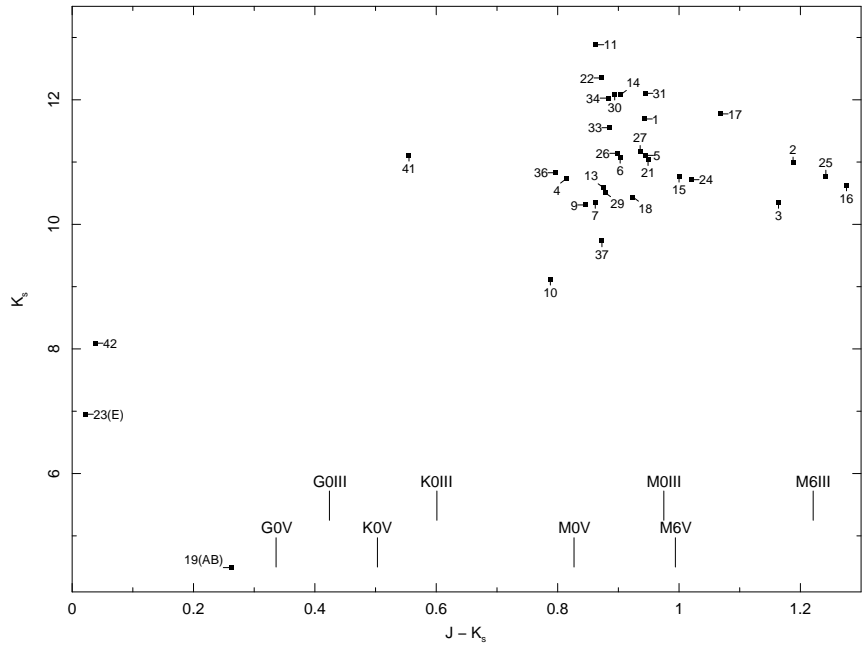}
\caption{Color-magnitude diagram of X-ray sources 
on ACIS-S3 having 2MASS identifications. Colors are
based on 2MASS photometry. Source numbers
correspond to  Table 2. Intrinsic colors are shown for 
late-type dwarfs and giants  from Bessell \& Brett (1988).
Source CXO 42 is the B5V binary star HD 37525. Source CXO 12 is
not shown and lies off the right edge of the plot
at  (J-K$_{s}$,K$_{s}$) = (2.33,14.04).}
\end{figure}

\clearpage

\begin{figure}
\figurenum{10}
\epsscale{1.0}
\includegraphics*[width=8.5cm,angle=-90]{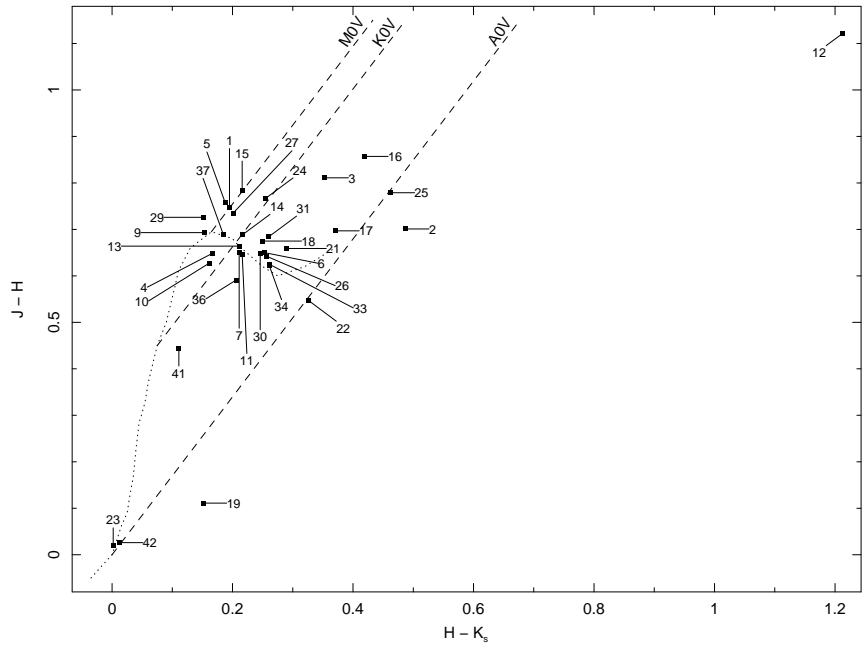}
\caption{Color-color diagram of X-ray sources 
on ACIS-S3 having 2MASS identifications. 
Colors are based on 2MASS photometry.
Source numbers correspond to  Table 2. The dotted line at 
left marks the unreddened main-sequence. 
The three sloping dashed lines show the loci
of  M0 V, K0 V, and A0 V stars 
based on intrinsic colors from Bessell \& Brett (1988) 
and the extinction law of Rieke \& Lebofsky (1985). }
\end{figure}

\clearpage

\begin{figure}
\figurenum{11}
\epsscale{1.0}
\includegraphics*[width=8.5cm,angle=-90]{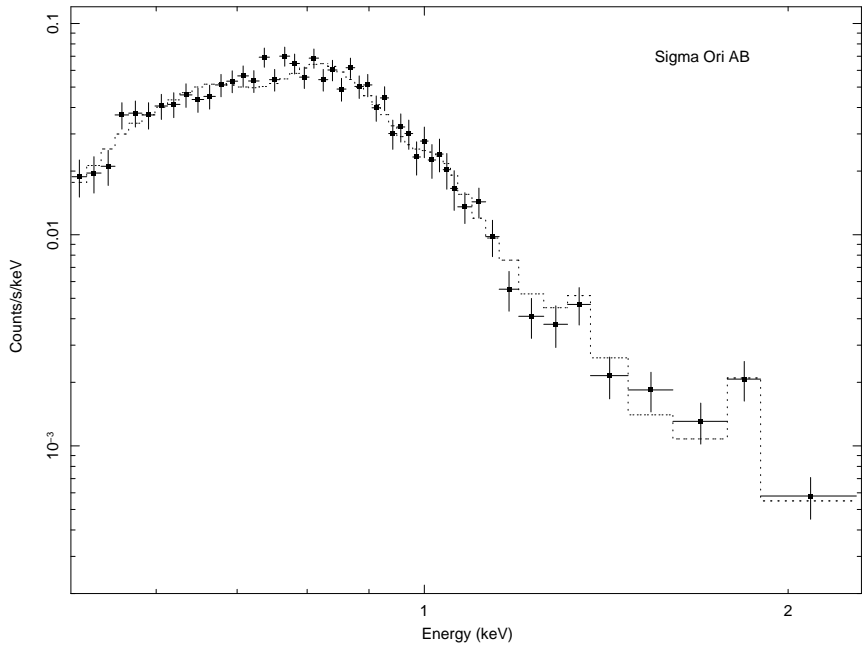}
\caption{Zero-order CCD spectrum of $\sigma$ Ori AB
rebinned to a minimum of 20 counts per bin. The dotted
line shows the best-fit 2T $apec$ model with fit parameters 
as listed in Table 3.}
\end{figure}

\clearpage

\begin{figure}
\figurenum{12}
\epsscale{1.0}
\includegraphics*[width=8.5cm,angle=-90]{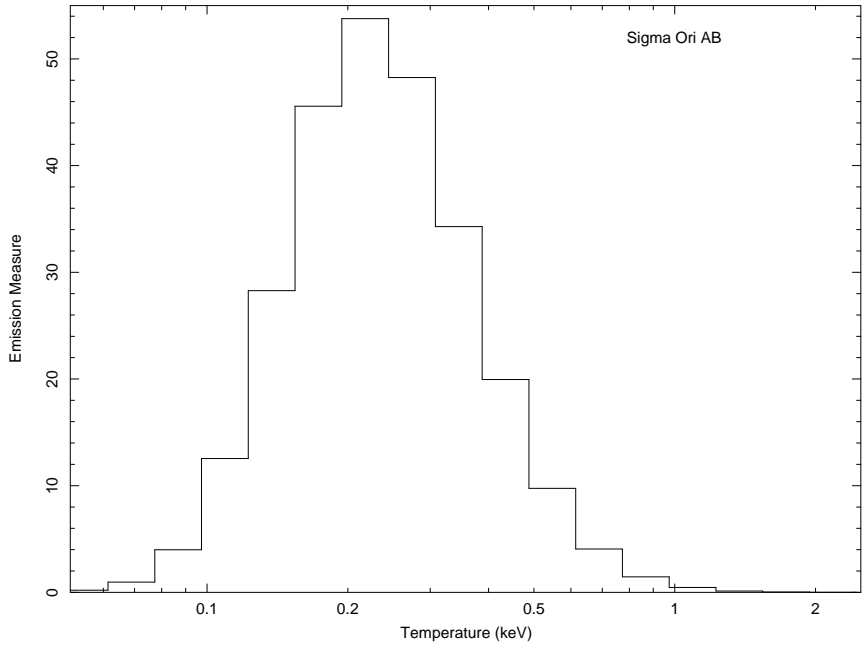}
\caption{Emission measure distribution  of $\sigma$ Ori AB
obtained from a best-fit of the zero-order CCD spectrum
with  the XSPEC model $c6pmekl$ using solar abundances
and hydrogen column density N$_{\rm H}$ = 3 $\times$ 10$^{20}$
cm$^{-2}$.}
\end{figure}

\clearpage

\begin{figure}
\figurenum{13}
\epsscale{1.0}
\includegraphics*[width=8.5cm,angle=-90]{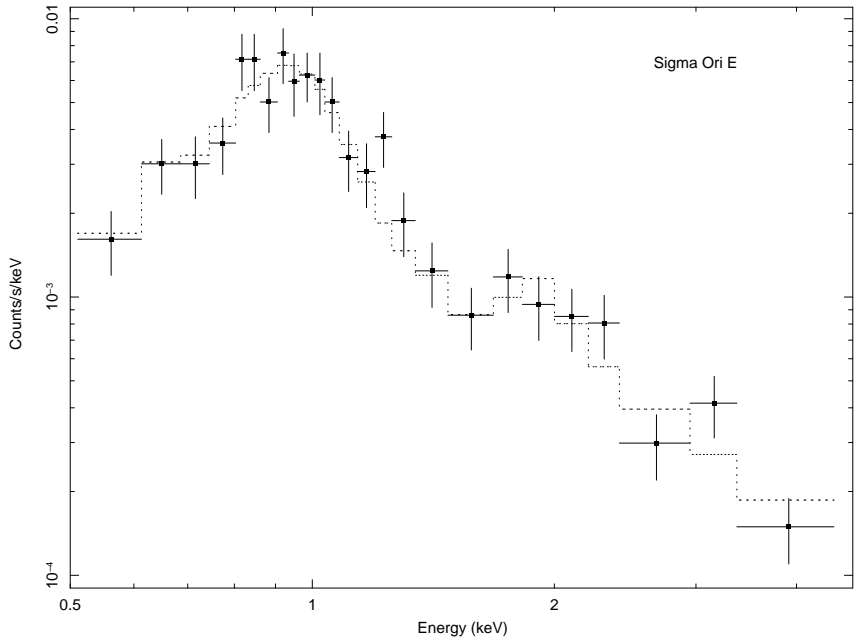}
\caption{Zero-order CCD spectrum of $\sigma$ Ori E
rebinned to a minimum of 15 counts per bin. The dotted
line shows the best-fit 3T $apec$ model with fit parameters 
as listed in Table 3 notes.}
\end{figure}

\clearpage

\begin{figure}
\figurenum{14}
\epsscale{1.0}
\includegraphics*[width=8.5cm,angle=-90]{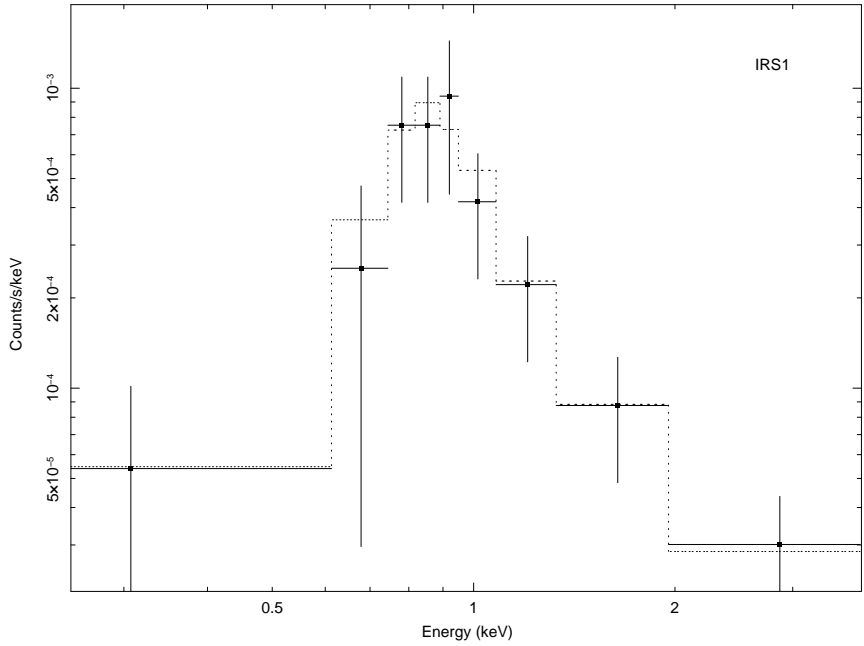}
\caption{ACIS-S3 CCD spectrum of $\sigma$ Ori IRS1 
(CXO 20) binned to  5 counts per bin.
The dotted line shows the 2T 
$apec$ model with fit parameters as listed in Table 3.}
\end{figure}

\clearpage

\begin{figure}
\figurenum{15}
\epsscale{1.0}
\includegraphics*[width=8.5cm,angle=-90]{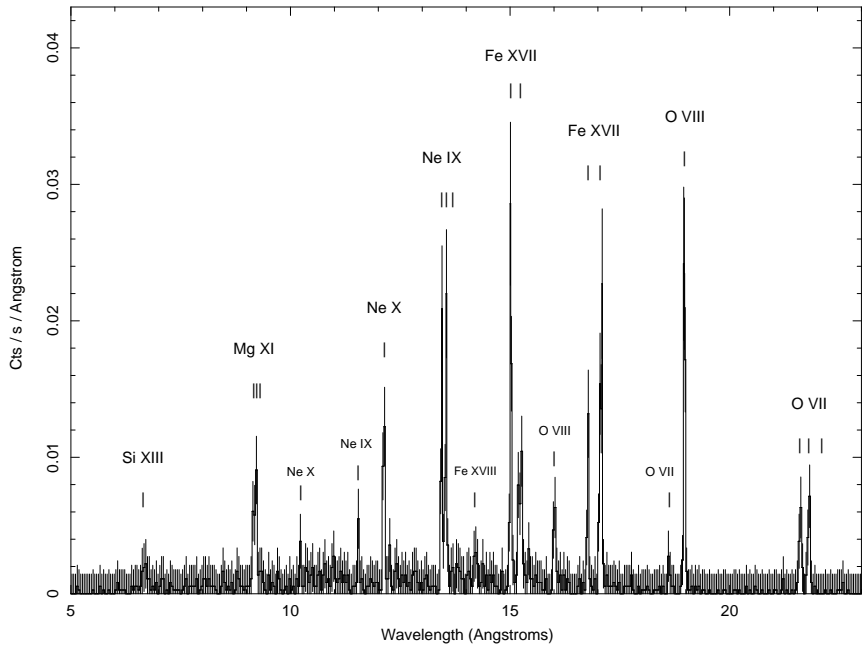}
\caption{HETG/MEG1 background-subtracted spectrum 
($+$1 and $-$1 orders combined; 1936 net counts)
of $\sigma$ Ori AB with prominent lines identified.}
\end{figure}

\clearpage

\begin{figure}
\figurenum{16}
\epsscale{1.0}
\includegraphics*[width=8.5cm,angle=-90]{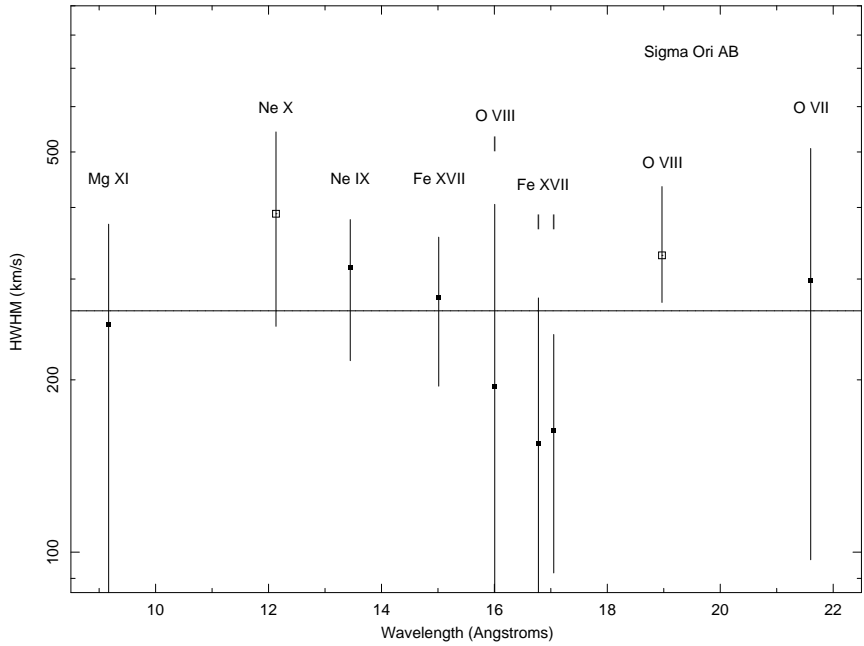}
\caption{Measured line widths of $\sigma$ Ori AB
expressed as half-width at half-maximum  
(1$\sigma$ error bars) versus line
laboratory wavelength using data from Table 4.
The two lines shown with open squares are blends
and their HWHM values are likely overestimated.
The solid horizontal line is the mean  value
$\overline{\rm HWHM}$ = 264 km s$^{-1}$.}
\end{figure}

\begin{figure}
\figurenum{17}
\epsscale{1.0}
\includegraphics*[width=8.5cm,angle=-90]{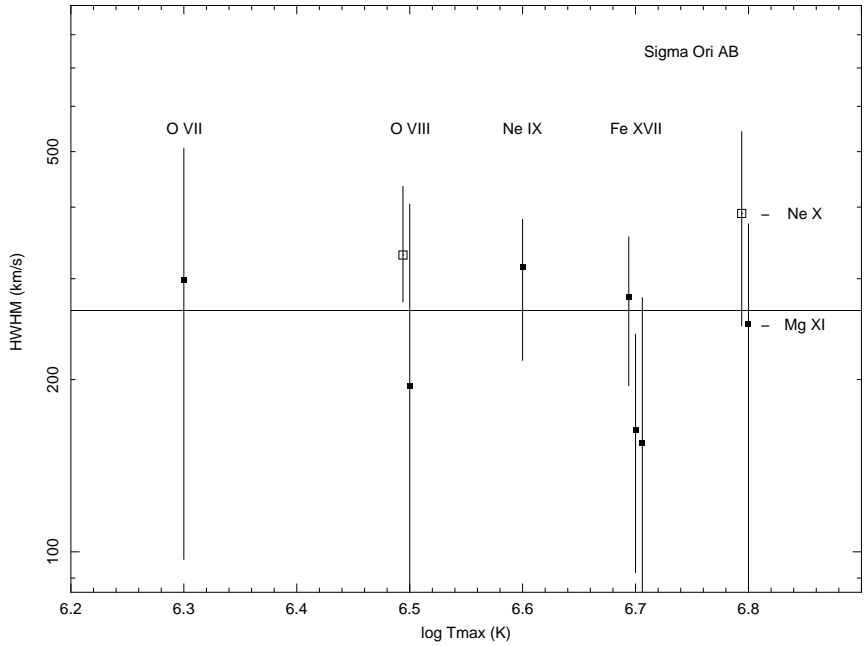}
\caption{Measured line widths of $\sigma$ Ori AB
expressed as half-width at half-maximum 
(1$\sigma$ error bars) versus maximum
line power temperature T$_{max}$ using data from
Table 4. The plotted values of T$_{max}$ for
O VIII, Fe XVII, and Ne X have been shifted 
slightly from the nominal values given in Table 4
to avoid error bar overlap.
The two lines shown with open squares are blends
and their HWHM values are likely overestimated.
The solid horizontal line is the mean value
$\overline{\rm HWHM}$ = 264 km s$^{-1}$.}
\end{figure}

\clearpage

\begin{figure}
\figurenum{18}
\epsscale{1.0}
\includegraphics*[width=8.5cm,angle=-90]{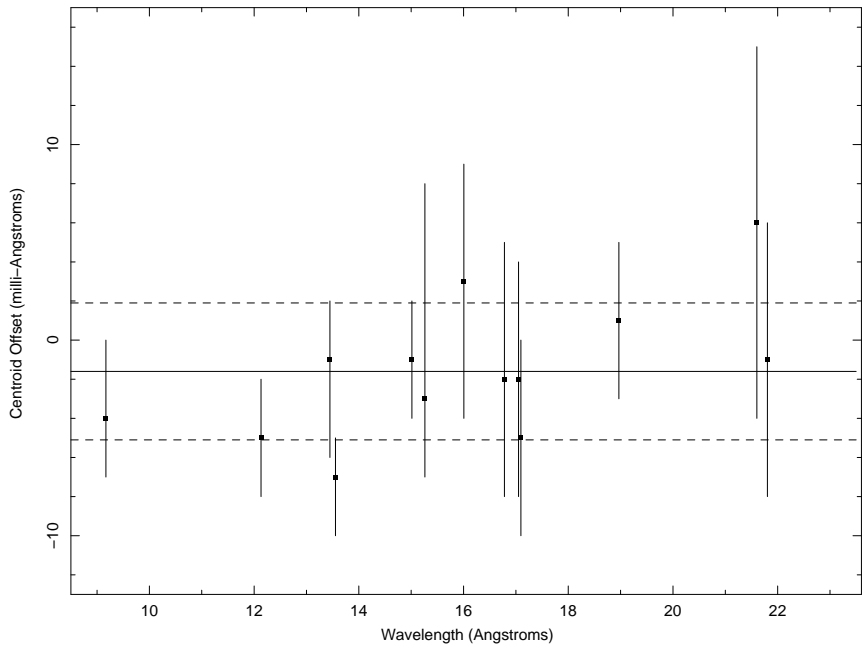}
\caption{Plot of the centroid offset  between the measured
and laboratory line centroid of prominent lines in the MEG1
spectrum, using   the offset values   
$\Delta$$\lambda$ = $\lambda_{obs}$ $-$
$\lambda_{lab}$  from Table 4. Errors bars are 1$\sigma$.
The mean offset (solid line) and $\pm$1$\sigma$ uncertainty
(dashed lines) for 13 measurements are $-$1.6 ($\pm$3.5) m\AA~.
}
\end{figure}

\clearpage

\begin{figure}
\figurenum{19}
\epsscale{1.0}
\includegraphics*[width=8.5cm,angle=-90]{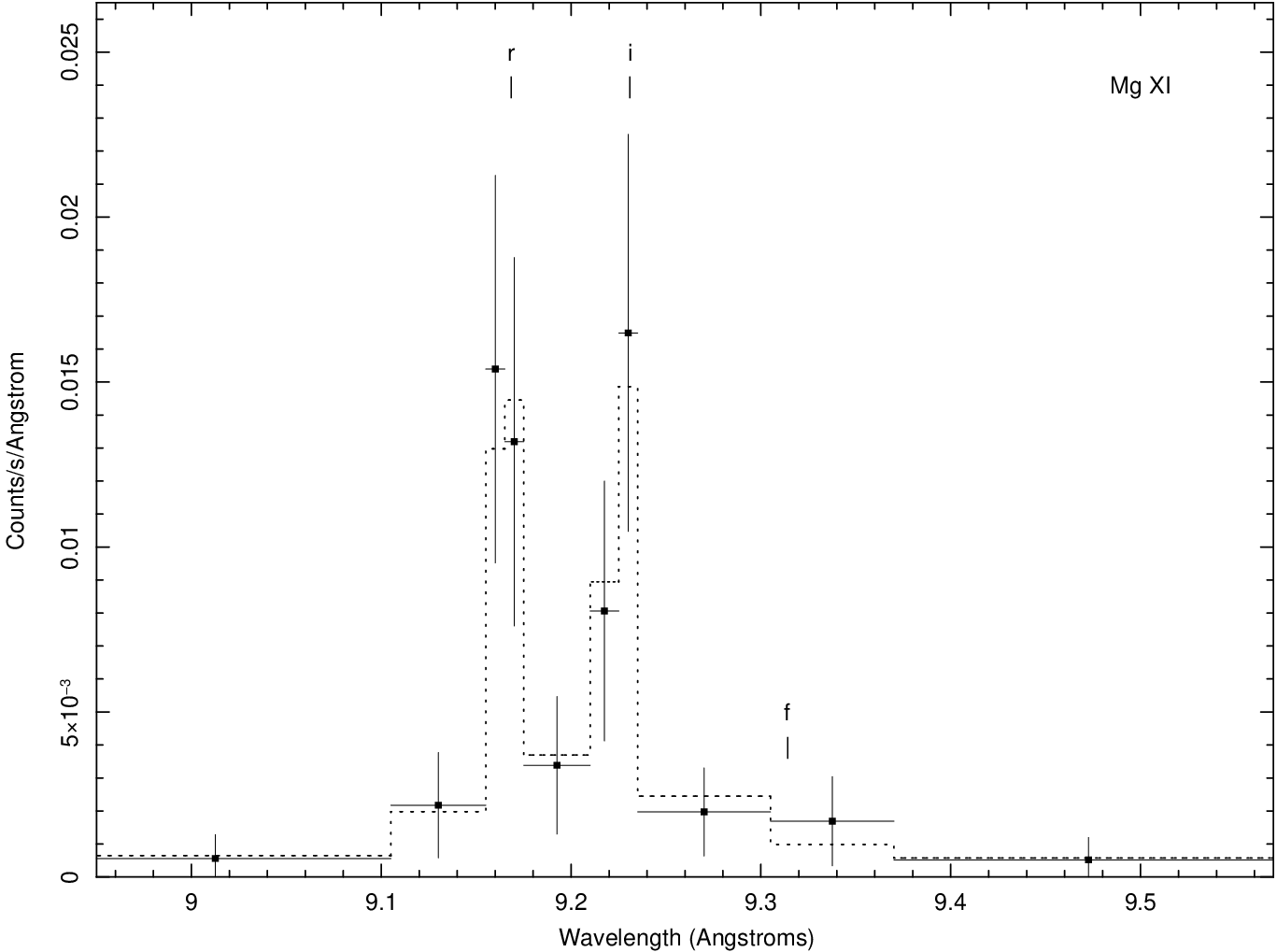}
\caption{Histogram plot of the MEG1 spectrum ($+$1 and $-$1 orders combined)
of $\sigma$ Ori AB near the Mg XI He-like triplet, binned to a minimum
of 10 counts per bin. Solid squares are data points and the dotted 
line is a Gaussian fit with weak power-law continuum. 
The resonance (r) and intercombination (i) lines
are detected but the forbidden (f) line is not.}
\end{figure}

\clearpage

\begin{figure}
\figurenum{20}
\epsscale{1.0}
\includegraphics*[width=8.5cm,angle=-90]{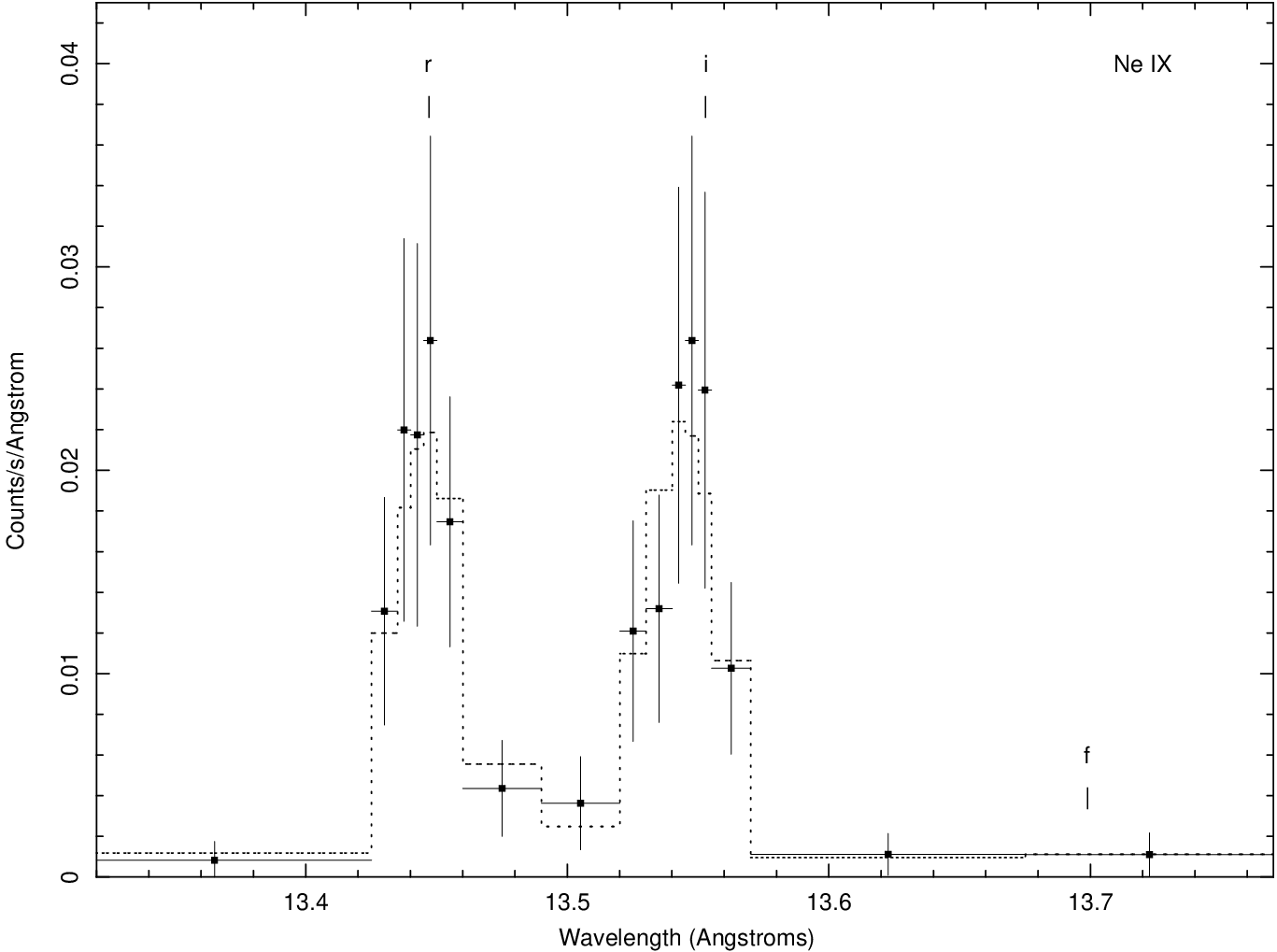}
\caption{Histogram plot of the MEG1 spectrum ($+$1 and $-$1 orders combined)
of $\sigma$ Ori AB near the Ne IX He-like triplet, binned to a minimum
of 10 counts per bin.  Solid squares are data points and the dotted
line is a Gaussian fit with weak power-law continuum. 
The resonance (r) and intercombination (i) lines
are detected but the forbidden (f) line is not.}
\end{figure}

\clearpage

\begin{figure}
\figurenum{21}
\epsscale{1.0}
\includegraphics*[width=8.5cm,angle=-90]{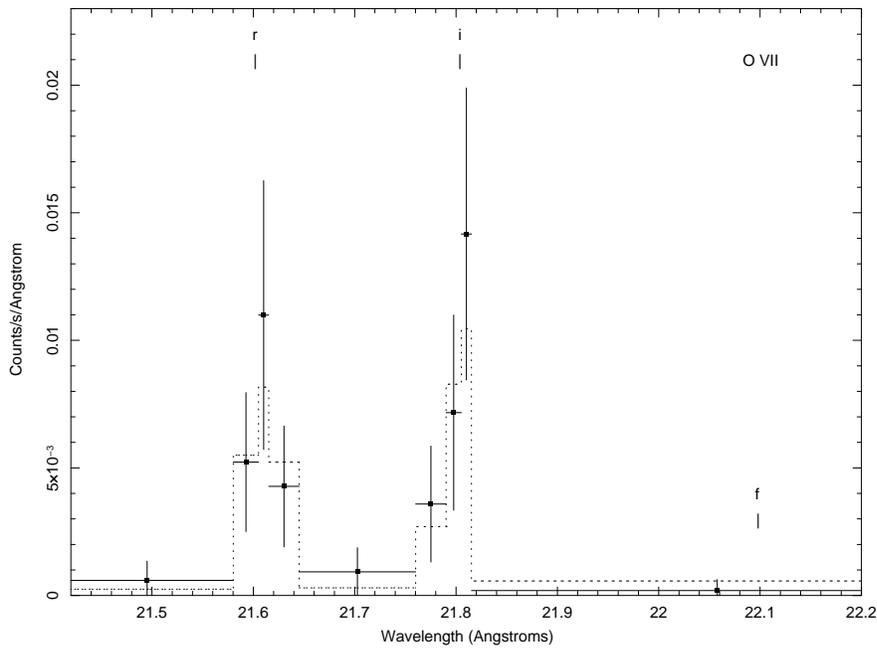}
\caption{Histogram plot of the MEG1 spectrum ($+$1 and $-$1 orders combined)
of $\sigma$ Ori AB near the O VII He-like triplet, binned to a minimum
of 10 counts per bin. Solid squares are data points and the dotted
line is a Gaussian fit with weak power-law continuum.
The resonance (r) and intercombination (i) lines
are detected but the forbidden (f) line is not.}
\end{figure}

\clearpage

\begin{figure}
\figurenum{22}
\epsscale{1.0}
\includegraphics*[width=8.5cm,angle=-90]{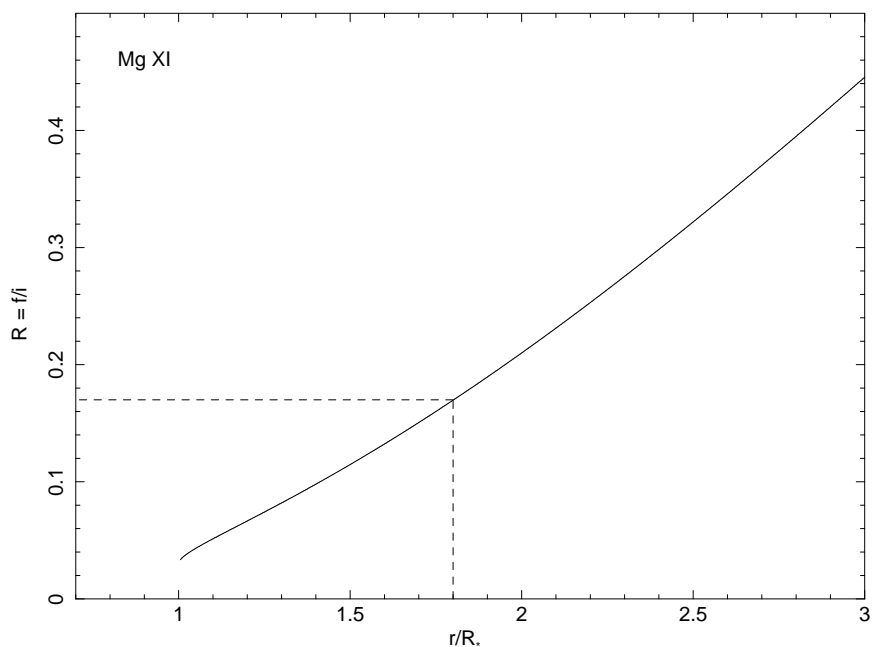}
\caption{The ratio of forbidden-to-intercombination Mg XI
line fluxes for $\sigma$ Ori AB as a function of distance
from the center of the star. The horizontal axis expresses
the distance in units of the stellar radius, where
R$_{*}$ = 9 R$_{\odot}$. The solid line shows the predicted
$R$ = f/i flux ratio based on a TLUSTY atmospheric model 
with T$_{eff}$ = 32,500 K (see text). The vertical dashed line
marks the 90\% upper limit on the line formation radius
r$_{fi}$ $<$ 1.8 R$_{*}$  (referenced to the center of the star)
based on the limit $R$ $<$ 0.17 determined from
spectral fits (Table 5). 
}

\end{figure}

\clearpage

\end{document}